\documentclass[]{bmcart}

\usepackage{amsthm,amsmath}
\RequirePackage{natbib}
\usepackage[utf8]{inputenc} %unicode support
\usepackage{graphicx}
\usepackage{soul}
\usepackage{footmisc}

\startlocaldefs

\usepackage[colorinlistoftodos]{todonotes}
\usepackage{url}

\definecolor{amethyst}{rgb}{0.6, 0.4, 0.8}

\newcommand\man[1]{{#1}}
\newcommand\revis[1]{{#1}}

\endlocaldefs

\begin{document}

%%% Start of article front matter
\begin{frontmatter}

\begin{fmbox}
\dochead{Research}


% \title{Swinging in the States:\\Does disinformation on Twitter mirror\\ the U.S. presidential election system?}
\title{Online disinformation in the 2020 U.S. Election: swing vs. safe states%Through the Lens of the Electoral System
}\thanks{Extended version of `Swinging in the states: Does disinformation on Twitter mirror the U.S. presidential election system?' WWW (Companion Volume) 2023: 1395-1403}

%%%%%%%%%%%%%%%%%%%%%%%%%%%%%%%%%%%%%%%%%%%%%%
%%                                          %%
%% Enter the authors here                   %%
%%                                          %%
%% Specify information, if available,       %%
%% in the form:                             %%
%%   <key>={<id1>,<id2>}                    %%
%%   <key>=                                 %%
%% Comment or delete the keys which are     %%
%% not used. Repeat \author command as much %%
%% as required.                             %%
%%                                          %%
%%%%%%%%%%%%%%%%%%%%%%%%%%%%%%%%%%%%%%%%%%%%%%

\author[
   addressref={imt,cnr},                   % id's of addresses, e.g. {aff1,aff2}
   corref={imt},                       % id of corresponding address, if any
   noteref={n1},                        % id's of article notes, if any
   email={manuel.pratelli@imtlucca.it}   % email address
]{\inits{MP}\fnm{Manuel} \snm{Pratelli}}
\author[
   addressref={cnr, imt},                
   %corref={cnr},                       
   noteref={n1},                        
   email={marinella.petrocchi@iit.cnr.it}  
]{\inits{MPe}\fnm{Marinella} \snm{Petrocchi}}
\author[
   addressref={cref, imt},                
   %corref={cref},                       
   noteref={n1},                        
   email={fabio.saracco@cref.it}  
]{\inits{FS}\fnm{Fabio} \snm{Saracco}}
\author[
   addressref={imt},                
   %corref={imt},                       
   noteref={n1},                        
   email={rocco.denicola@imtlucca.it}  
]{\inits{RDN}\fnm{Rocco} \snm{De Nicola}}

%%%%%%%%%%%%%%%%%%%%%%%%%%%%%%%%%%%%%%%%%%%%%%
%%                                          %%
%% Enter the authors' addresses here        %%
%%                                          %%
%% Repeat \address commands as much as      %%
%% required.                                %%
%%                                          %%
%%%%%%%%%%%%%%%%%%%%%%%%%%%%%%%%%%%%%%%%%%%%%%

\address[id=imt]{%                           % unique id
  \orgname{IMT Scuola Alti Studi Lucca}, % university, etc
  \street{Piazza San Francesco 19},                     %
  \postcode{55100}                                % post or zip code
  \city{Lucca},                              % city
  \cny{Italy}                                    % country
}
\address[id=cref]{%                           % unique id
  \orgname{`Enrico Fermi' Research Center}, % university, etc
  \street{via Panisperna 89A},                     %
  \postcode{00184}                                % post or zip code
  \city{Rome},                              % city
  \cny{Italy}                                    % country
}
\address[id=cnr]{%                           % unique id
  \orgname{Istituto di Informatica e Telematica CNR}, % university, etc
  \street{via G. Moruzzi 1},                     %
  \postcode{56124}                                % post or zip code
  \city{Pisa},                              % city
  \cny{Italy}                                    % country
}

%%%%%%%%%%%%%%%%%%%%%%%%%%%%%%%%%%%%%%%%%%%%%%
%%                                          %%
%% Enter short notes here                   %%
%%                                          %%
%% Short notes will be after addresses      %%
%% on first page.                           %%
%%                                          %%
%%%%%%%%%%%%%%%%%%%%%%%%%%%%%%%%%%%%%%%%%%%%%%

\begin{artnotes}
%\note{Sample of title note}     % note to the article
\note[id=n1]{Equal contributor} % note, connected to author
\end{artnotes}

\end{fmbox}% comment this for two column layout

%%%%%%%%%%%%%%%%%%%%%%%%%%%%%%%%%%%%%%%%%%%%%%
%%                                          %%
%% The Abstract begins here                 %%
%%                                          %%
%% Please refer to the Instructions for     %%
%% authors on http://www.biomedcentral.com  %%
%% and include the section headings         %%
%% accordingly for your article type.       %%
%%                                          %%
%%%%%%%%%%%%%%%%%%%%%%%%%%%%%%%%%%%%%%%%%%%%%%

\begin{abstractbox}

\begin{abstract} % abstract

  % aggiustato il testo (sotto)
  %Our intuition is that election debate may cause more traffic on Twitter-and probably be more plagued by misinformation-when associated with swing states. The results mostly confirm the intuition.
  %About 88\% of the entire traffic can be associated with swing states, and links to non-trustworthy news are shared  far more in swing-related traffic than the same type of news in safe-related traffic.
  %Considering traffic origin instead, non-trustworthy tweets generated by automated accounts, so-called social bots, are mostly associated with swing states.
  %\man{Furthermore, two main communities emerge, one displaying a purely Republican affiliation, the other one including accounts with mixed political orientation. We observe that the vast majority of the disinformation flow is concentrated in the first community.} 
  %Our work sheds light on the role an electoral system plays in the evolution of online debates, with, in the spotlight, disinformation and social bots.
\man{For U.S. presidential elections, most states use the so-called winner-take-all system, in which the state's presidential electors are awarded to the winning political party in the state after a popular vote phase, regardless of the actual margin of victory. Therefore, election campaigns are especially intense in states where there is no clear direction on which party will be the winning party. These states are often referred to as {\it swing states}}.
To measure the impact of such an election law on the campaigns, we analyze the Twitter activity surrounding the 2020 US preelection debate, with a particular focus on the spread of disinformation. We find that about 88\% of the online traffic was associated with swing states. In addition, the sharing of links to unreliable news sources is significantly more prevalent in tweets associated with swing states: in this case, untrustworthy tweets are predominantly generated by automated accounts.
Furthermore, we observe that the debate is mostly led by two main communities, one with a predominantly Republican affiliation and the other with accounts of different political orientations. Most of the disinformation comes from the former.

\end{abstract}

%%%%%%%%%%%%%%%%%%%%%%%%%%%%%%%%%%%%%%%%%%%%%%
%%                                          %%
%% The keywords begin here                  %%
%%                                          %%
%% Put each keyword in separate \kwd{}.     %%
%%                                          %%
%%%%%%%%%%%%%%%%%%%%%%%%%%%%%%%%%%%%%%%%%%%%%%


% MSC classifications codes, if any
%\begin{keyword}[class=AMS]
%\kwd[Primary ]{}
%\kwd{}
%\kwd[; secondary ]{}
%\end{keyword}

\end{abstractbox}
%
%\end{fmbox}% uncomment this for twcolumn layout

\end{frontmatter}

%%%%%%%%%%%%%%%%%%%%%%%%%%%%%%%%%%%%%%%%%%%%%%
%%                                          %%
%% The Main Body begins here                %%
%%                                          %%
%% Please refer to the instructions for     %%
%% authors on:                              %%
%% http://www.biomedcentral.com/info/authors%%
%% and include the section headings         %%
%% accordingly for your article type.       %%
%%                                          %%
%% See the Results and Discussion section   %%
%% for details on how to create sub-sections%%
%%                                          %%
%% use \cite{...} to cite references        %%
%%  \cite{koon} and                         %%
%%  \cite{oreg,khar,zvai,xjon,schn,pond}    %%
%%  \nocite{smith,marg,hunn,advi,koha,mouse}%%
%%                                          %%
%%%%%%%%%%%%%%%%%%%%%%%%%%%%%%%%%%%%%%%%%%%%%%

%%%%%%%%%%%%%%%%%%%%%%%%% start of article main body
% <put your article body there>

%%%%%%%%%%%%%%%%
%% Background %%
%%

\section{Introduction}
\man{}

%Abbomba :)

%FABIO aggiungo un REF ad un articolo su Nature che avevo trovato 

The proliferation of online disinformation has emerged as a pressing concern, akin to a modern-day plague. As major events unfold, be they elections, public health crises, or geopolitical shifts, disinformation often takes center stage, sowing confusion and distrust among the public\footnote{\url{https://www.un.org/en/countering-disinformation} All urls were last accessed on January 26, 2024.},  \man{with potentially resounding offline consequences. To give a glaring example, on 3 September 2021, Jacob Anthony Angeli Chansley was sentenced to 41 months in prison for obstruction of justice. Chansley, also known by various nicknames such as ``QAnon Shaman'', participated with other far-right activists in the attack on the United States Capitol on January 6, with the intention of disrupting the certification of election results. He was convinced by online disinformation campaigns about fraud against former President Donald Trump in the election\footnote{\man{\url{https://en.wikipedia.org/wiki/Jacob_Chansley}}. } and a known conspiracy theorist~\cite{tollefson2021trump}. This egregious news episode is just the tip of the iceberg of a series of consequences that the proliferation of disinformation online has on society around election time. A nationwide survey in the U.S. after the 2018 midterm elections found
that trust in the electoral system dropped significantly after Republican supporters spread unsubstantiated rumors of fraud online,
driving voters away from politics, despite fact-checking efforts to disprove such rumors \cite{Berlinski2023}.}

Numerous scholars have delved into the U.S. 2016 and 2020 presidential elections, seeking to gauge the extent to which digital disinformation influenced Trump's victory/defeat, yet a definitive answer remains elusive. For example, the study by Georgacopoulos et al.~\cite{Georgacopoulos2020how} reveals that in the three months leading up to the 2016 election, fake news supporting Trump was shared on Facebook nearly four times more than the eight million fake news items supporting Clinton.

Examining more than 170 million tweets exchanged on Twitter in the five months leading up to the same election, Bovet and Makse~\cite{Bovet2019influence} found that trustworthy news stories overwhelmingly came from journalistic sources and verified Twitter accounts. In contrast, conspiracy theories, fake news, and highly partisan news largely originated from unofficial Twitter clients, posted by unknown users who often disappeared from the platform, or through automated accounts commonly referred to as social bots.
Shao et al.~\cite{shao2018anatomy} also highlighted the role of Twitter bots, showing how these bots were primarily responsible for the early spread of disinformation by engaging influential accounts through mentions and replies.

The evolutionary adaptation of bots, characterized by their increasing ability to evade detection techniques, is well documented~\cite{Ferrara2016rise,Cresci2017paradigm}. Luceri et al.~\cite{luceri2019evolution}, for example, found that from the 2016 presidential election to the 2018 midterm elections, political discussion bots evolved to the point where they became increasingly indistinguishable from humans.
Analyzing the manipulation of online narratives around the 2020 U.S. election, Ferrara et al.~\cite{Ferrara2020characterizing} found that a relatively small number of automated accounts managed to generate traffic spikes in election discourse comparable in scale to human users, who significantly outnumbered them.

Our study focuses on the Twitter debate during the week leading up to November 4, 2020. Like previous research, we examine the flow of disinformation and the infiltration of bots into this discourse. However, our work uniquely focuses on two specific aspects of the U.S. presidential election:  the presence of swing and safe states and the winner-take-all system. Recent literature comparing online political debates across countries highlights how different electoral systems lead to different structural properties within online social networks~\cite{Bright2018,Urman2020,VanVliet2021,Praet2021}.

The term `swing' refers to states where a landslide victory for either Republicans or Democrats is uncertain, owing to the lack of a clear %historical 
voting orientation. 
%OLD version
%In contrast, a state is classified as `safe' if its residents have consistently elected representatives from the same political party in recent history.
\man{In contrast, a state is deemed `safe' when the electoral races are not competitive and are unlikely to be closely contested. Competitiveness is determined by several factors, including the political composition of the state and its counties, the prevailing local and national political climate, and insights from interviews with campaign experts\footnote{\url{https://www.cookpolitical.com/ratings/presidential-race-ratings}}. Therefore, it is important to note that the status of swing and safe states is not fixed. Major swing states may become reliably safe Republican or Democratic states over time, while traditionally solid red or blue states may move into the swing state category. Changing demographics and political realignments within specific regions or demographic groups often drive these shifts\footnote{\url{https://www.maynoothuniversity.ie/research/spotlight-research/10-swing-states-will-decide-us-presidential-election}}.}

% Shifting demographics and the political realignment of certain regions or social or demographic groups can drive these changes

With the exception of Maine and Nebraska, all U.S. states utilize the winner-take-all voting method. Each state has a varying number of presidential electors, determined in part by its population. Following a popular vote, each state allocates its presidential electors based on the candidate with the most votes, \man{due to} %courtesy of 
the winner-take-all system.
A major criticism of this system is that it incentivizes presidential candidates to focus their campaigns on a select few swing states, as they hold the key to victory\footnote{\url{https://www.jstor.org/stable/j.ctt1npxbp}}. In particular, certain battleground states, such as Florida, traditionally a swing state with a substantial population and a large allocation of presidential electors\footnote{\url{https://edition.cnn.com/election/2020/results/state/florida}}, have been subjected to more intense electoral campaigns.
Transferring this critique to the realm of Twitter, our paper poses and answers a central question: Could it be that the Twitter discourse leading up to the 2020 U.S. presidential election mirrors the electoral system, specifically the distinction between swing and safe states?

Specifically referring to disinformation flows,

\begin{itemize}

    \item Is there a difference in the frequency of tweets containing links to dubious or unreliable news when they are associated with swing states or safe states during the 2020 pre-election season? Is this difference in frequency also related to the political orientation of the account?
    
    \item Does the prevalence of automated accounts in online pre-election political debates differ depending on whether the discussion focuses on swing states or safe states? If so, is the difference also related to the political orientation of the account? 
\end{itemize}

To perform the analysis, we collected Twitter data using keyword searches, specifically pairing candidate and state names.
We then processed the data and the users who created and shared it as follows. First, we extracted links to news stories in the tweets and associated those stories with a level of trustworthiness. Second, we classified the users as bots or not. Third, we extracted the main discursive communities and their political orientation, which we used to (i) filter out irrelevant data from the entire dataset, specifically users who were not interested in the political narrative, and (ii) gain insight into the specific political leanings of the accounts.

\subsection{Contributions} Our main contributions are:
\begin{itemize}
    \item We provide a fine-grained characterization of the Twitter traffic about the 2020 U.S. presidential election, in the week leading up to Election Day, adopting a multidisciplinary approach including complex network analysis, to identify non-trivial communities of users and their political leanings, 
    %interested in the political narrative)}, 
    %(i.e., entropy-based null-model), 
    artificial intelligence (to classify users as bots or not), 
    %(i.e., Botometer tool) 
    and human-based annotation (to classify news sources  as trustworthy or not).
    %(i.e., NewsGuard's expert annotations).

    %provato a rifrasare / migliorare sotto
    %\item To the best of our knowledge, this is the first paper that addresses the problem of understanding whether, in the Twittersphere, \man{}\man{during the U.S. 2020 pre-election period, the online debate was centered more on tweets about swing than safe states.}%, the U.S. 2020 pre-election debate was polarized more in tweets about  swing states rather than safe states.

    \item To the best of our knowledge, this is the first paper that investigates the links between the U.S. presidential electoral system and the online debate about the election, focusing on automated accounts, the diffusion of low-credible news, and employing a sophisticated network-based approach to identify the specific political leanings of the users participating in the debate.
    
    %, by spreading more disinformation content associated to the former than to the latter. 
    
    %riformulata sotto
    %\item We provide evidence of an alignment between the real electoral mechanism -which often favors more intense campaigning in certain 
    %\man{locations (\textbf{MEGLIO CAMBIARE LOCATION})}
    %\man{states (i.e., in swing)}- and the online electoral debate. Indeed, the 2020 election-related traffic focuses, for the vast majority, on tweets about swing states; swing state-related debate sees a higher concentration of links to non-trustworthy news sites. 
    % \man{Furthermore, considering only the traffic generated by classified users,} 
   %The majority of disinformation content associated with swing states 
    %\man{and (ii) Republican supporters}
%is posted and retweeted by automated accounts.

    \item 
    We provide compelling evidence of a correlation between the actual electoral mechanism, which tends to prioritize intense campaigning in swing states, and the online electoral debate. Indeed, we observe that a significant portion of the 2020 election-related online traffic revolves around tweets focused on swing states. Furthermore, the discourse surrounding swing states exhibits a higher concentration of links leading to untrustworthy news sites. Importantly, most of the disinformation content associated with swing states (and Republican supporters) originates from automated accounts, indicating their significant role in spreading such content. 
\end{itemize}

% \subsection{RQs}

% Considering discussion inside the main emergent (politically oriented discursive) communities and related to swing/
% safe states:

% \begin{itemize}
%     \item [RQ1] Nelle discussioni si formano delle community associabili a correnti politiche ben definite?
%     \item [RQ2] Are there significant differences in terms of misinformation? Polarization?
%     \item [RQ3] Are there significant differences in terms of bot presence?
%     \item [RQ4] Are there correlations between the presence of bots and the spread of low credibility contents supporting misinformation? Polarization?
% \end{itemize}

\subsection{Results}
The experiments conducted in this work
%guided by the defined research questions, 
led us to the following results:
\begin{itemize}

    \item Tweets associated with swing states account for about 88\% of the whole traffic.  % commentato 
    \man{As a rough measure, the population of the swing states in the dataset represents 66\% of the population of the states in our dataset. In this sense, the swing states have received more attention than would have been expected based solely on the number of electoral votes they represent.} 

    \item Two main user communities emerge from the data: a homogeneous one, consisting of Republican supporters (hereafter referred to as {\sc Rep}), and a mixed one, comprising journalists as well as both Republican and Democratic supporters (hereafter referred to as {\sc Rep-Dem-Journ}).
    % \item Dai dati sono emerse \textbf{due principali community} di utenti: una costituita principalmente da sostenitori repubblicani (di qui in avanti chiamata "rep") l'altra di sostenitori misti repubblicani, democratici e giornalisti ("rep\_dem\_journ");(vedi caratterizzazione community in Tab \ref{tab:community_characterization})

    \item More than 90\% of links to news from untrustworthy publishers are concentrated in the {\sc Rep} community. Each of these links is shared an average of 57 times, a significantly higher number than the average number of shares in the {\sc Rep-Dem-Journ} community (7).

    \item The percentage of tweets with URLs pointing to news from untrustworthy publishers is consistently higher for swing states in all communities.%than for safe states (both over the whole dataset, {\sc Rep} and {\sc Rep-Dem-Journ}).}

    % \item The percentage of links to non-trustworthy news sites is higher in tweets associated to swing states - $\sim$23.5\% - than in those associated to safe states - $\sim$18\%. \mari{However, the gap is not such that a claim can be made about a net majority of disinformation flow  in traffic associated with swing rather than safe states.}
    % \item The percentage of trustworthy links  is higher in tweets about safe states - $\sim$51\% ($\sim$30\% in tweets about swings).
    
    \item Tweets associated with safe states have a higher concentration of URLs pointing to news with trustworthy publishers. Tweets associated with swing states have a higher concentration of URLs pointing to news with untrustworthy publishers.% in all communities.

    \item Of the total number of tweets associated with swing states and containing untrustworthy URLs, 74\% of these are posted or retweeted by accounts classified as bots. 
    %Accounts classified as bots in {\sc Rep} generate non-trustworthy tweets that are associated more with swing states than safe states ($\sim$74\% swing {\it vs} $\sim$68\% safe). 

\end{itemize}

\subsection{Originality}
This work is neither the first nor the last to address the impact of real-world events on virtual ones, and vice versa. \man{A brief review on the relationships between electoral politics and social media will be presented later in this article.} Work of Howard et al. in~\cite{howard2018social} examines tweets from authors who left some evidence of their physical location in the period leading up to the 2016 U.S. presidential election. The analysis reveals a high concentration of polarized news in tweets associated to  swing states with a significant number of presidential electors.
In addition to differences in years (2016 versus 2020) and differences in data collection methods (hashtags versus general keywords), our study differs from the work of Howard et al.~\cite{howard2018social} in some important ways. First, our analysis includes an evaluation of automated accounts, and the classification of news sources is based on the annotations of expert journalists.

However, the primary distinction of our study lies in the rigorous filtering process applied to our dataset. This process employs advanced statistical methods specifically tailored for the analysis of complex networks, making them well suited for the study of interactions within social networks. For a complete understanding of these methodologies, the reader is referred to Sections \ref{sec:rw} and \ref{sssec:DisCo}.
%an entropy-based null model for bipartite networks known as the Bipartite Configuration Model BiCM~\cite{Cimini2018a}.
\man{Using this filtering process allows us to %achieve two primary objectives: (i) to extract political discussions from the overall discourse, and (ii) to 
gain insight into the political affiliations of users participating in these discussions (i.e., which political party users tend to be more closely associated with).}

\subsection{Change in Twitter property and the advent of Twitter/X}
In late October 2022, Twitter, Inc., the American social media company, underwent a significant transformation when it became the property of Elon Musk\footnote{\url{https://www.nytimes.com/2022/10/27/technology/elon-musk-twitter-deal-complete.html}}. This transition ushered in a series of radical changes and reforms that embraced both managerial and technical aspects.

One pivotal development of note for the scientific community was the discontinuation of Twitter's free API tier by February 2023, to be replaced with a `basic paid tier'\footnote{\url{https://twitter.com/XDevelopers/status/1621026986784337922}}. For researchers and developers, this change meant that Twitter content was no longer available for research purposes without subscribing to a significantly different paid plan. It also presented a challenge in terms of rehydrating the datasets currently in use.
Despite the fact that our dataset was collected during a period of free access (from October 27 to November 3, 2020), the policy appears to be unchanged at the time of revising this manuscript~\footnote{\url{https://developer.twitter.com/en/developer-terms/policy#4-e}}. Thus, we acknowledge the potential obstacles to the reproducibility of the experiments presented here.

However, we maintain that our methodology remains highly adaptable to other online social networks. It relies on two core principles: (i) the analysis of account activity related to the sharing of news source URLs, and (ii) the identification of discursive communities. Although extending the former to alternative social platforms is straightforward, the latter may present some challenges. In our case, we used the activity of verified users, a prominent group of content creators on Twitter~\cite{Becatti2019d}. \man{Even if not all social platforms offer such certification, when analyzing other platforms we can still focus on other categories of users that play a central role in shaping public discourse, such as influential users as defined in previous studies~\cite{Gonzalez2013}}.

\section{Related Work}
\label{sec:rw}

\paragraph{Electoral politics and social media}
\revis{The basic question in our article is whether the U.S. presidential election system, whose peculiarities make physical campaigning more vibrant in so-called swing states, mimics this aspect in online campaigning. 
From this perspective, we found it interesting to examine several works that have studied the relationship between electoral politics and the use of social media. 

In particular, one paper used opinion mining techniques to examine in real-time the correspondence between exit poll results and the opinions of Twitter users in the week leading up to the 2020 U.S. election \cite{belcastro2022voter}. In fact, it was possible to predict the president-elect in 10 of the 11 states considered to be swing states in 2020, even exceeding the percentages of the most recent physical exit polls. In this case, we could say that there was more than a campaign mimicry, there was a match precisely in the prediction of the winner.

Similarly, a series of articles focused on the influence that exposure to Facebook and Instagram feeds had on the voting decisions of U.S. citizens during the 2020 presidential election campaign\footnote{Social Media and the 2020 election: \url{https://www.princeton.edu/news/2023/07/28/social-media-polarization-and-2020-election-insights-spias-andrew-guess-and}}.

In a first paper \cite{guess2023feed}, nearly 45k users from the two platforms were recruited. One control group retained feed visualization settings dictated by the platforms’ algorithms, while the others had their settings altered, allowing users to see the most recent feeds.
Despite these changes, being fed content based on history rather than interests reportedly had no effect on the users' political attitudes and off-platform behaviors. 

A second paper was interested in the effects of viewing news shares on Facebook, again during the U.S. 2020 election period \cite{Guess2023resharing}. The removal of re-shares significantly reduced the amount of political news, including content from untrusted sources. Despite this, there were no changes in political polarization or individual-level political attitudes in the subgroup that did not see shared content. From these studies, it appears, as above, that online users were little influenced by the content proposed by the platform and generated by other users, i.e. their voting intention was not surprisingly changed.

 Many works on the relationship between electoral politics and social media focus on Europe, and in particular the European Parliament elections. The number of studies is probably due to the different facets that this particular election has in Europe, from the number of elected officials from each country, the presence of numerous local parties, and the doubt that campaigns are driven more by individual member state interests than by all as a community. In fact, according to experts, the European Parliament elections are experienced by European citizens as 27 different election campaigns, one for each member state\footnote{Make or break for the EU? Europeans vote in June with far right on the rise: \url{https://www.theguardian.com/world/2024/jan/03/make-or-break-for-the-eu-europeans-vote-in-june-with-far-right-on-the-rise}}.

One of the study's findings is similar to ours: the more citizens the candidate is expected to represent, the more activity the candidate has on the social network. Thus, when the physical campaign becomes heavy and complex to manage, the campaign activity is changed online. 
However, the social activity of the Member of the European Parliament -or MEP- candidates is limited to the election period, indicating that social networks are not used to cement a relationship with the electorate, but rather for the sole purpose of garnering votes. Some studies justify the use of social networks by candidates as a means of broadcasting only on the grounds that interaction with users leads to insults and harassment. This was the result of an analysis in \cite{Theocharis2016workman}, which found that the content of tweets directed at MEP candidates was often rude and harassing.

A continuation of the study in \cite{Daniel2016electoral}  is the one in \cite{Daniel2020reaching}, where the authors looked at outgoing members of the European Parliament who, after the 2014 elections, ran again as candidates in the same elections in 2019. 
The purpose of the work was to see whether the different candidates focused their online campaigns more on themselves or on the party they represented. One finding that emerged, compared to running a physical campaign, was that the relatively low-cost nature of social media allowed some politicians to simultaneously campaign as individuals and as \textit{party animals} in a way that analog campaigning could not.

Social campaigns for the 2019 European Parliament have been widely studied in the literature. 
The article collection in \cite{habler2021} analyzes how political parties in 12 member states used Facebook in the lead-up to elections. Again, the overall message is that social media was used to persuade the public to vote for the candidate, rather than using the platform to interact and mobilize voters. 

% \cite{Daniel2016electoral}
% twitter e EP, electoral connection

%\cite{Theocharis2016workman}  sempre elezioni parlamento europeo, come mai i politici non usano twitter piu' a lungo per campagne elettorali..?

This is also the conclusion of other work that has examined the relationship between social campaigning and electoral politics in the run-up to national elections in Europe. One example is the study in \cite{bright2020campaigning}, in which two electoral events in Britain, in 2015 and 2017, are considered. What comes out is that, once again, the tendency of candidates to use the social medium is to broadcast the program and make propaganda, and not to interact with the electorate, even with the idea of maintaining solid contact after the election.

}

\paragraph{Disinformation flows \man{in U.S. presidential elections}} 
% UNIRE SOPRA IN QUALCHE MODO :) Fabio: Dici? Anche così mi sembrava ok…
%\man{The 2020 U.S. presidential election has raised concerns about its integrity, shared both by the U.S. government and the general public.} 
In the introduction, we already cited some analyses on detecting online disinformation flows in the periods leading up to  the 2016 and 2020 U.S. presidential elections~\cite{Georgacopoulos2020how,Bovet2019influence, shao2018anatomy,Ferrara2020characterizing}. 
%Here we report a summary of other work that has examined online disinformation flows during major political events, both in the U.S. and elsewhere.
We can also cite~\cite{faris2017partisanship}, where the authors analyzed both mainstream and social media coverage of the 2016 U.S. presidential election. Their analysis revealed the asymmetric nature of the media landscape, with Twitter displaying a more partisan tendency. Donald Trump's campaign primarily emphasized immigration, whereas Hillary Clinton's coverage tended to emphasize various scandals. Right-wing media tended to favor pro-Trump outlets, while left-wing media focused on traditional objective journalism. 

As mentioned at the beginning of the article, Howard et al. in~\cite{howard2018social} conducted a study centered on analyzing tweets related to swing and safe states during the pre-election period of the 2016 U.S. presidential election. Their findings revealed a significant concentration of polarized news in tweets associated with swing states with a significant number of presidential electors. This work serves as a valuable precursor to our current study. It is important to note, however, that the election we examine differs from the one in their study. In addition, our research focuses on the behavior of social bots, and most importantly, we have refined our dataset by employing a process based on complex network analysis to filter out noise.

\paragraph{Statistical methods for the analysis of online social networks}

The recent literature regarding online social networks has progressively implemented more techniques based on network science, with the aim of distinguishing  non-trivial signals of social interactions from random noise. 

In particular, the implementation of entropy-based null-models (see the review by Cimini et al~\cite{Cimini2018a}) has opened up a variety of applications, providing a general and unbiased benchmark for the analysis of complex networks.
%More in details, such methods were implemented for the identification of discursive communities on Twitter, i.e. groups of accounts based on their retweeting activities~\cite{Becatti2019d,caldarelli2020role, Radicioni2021a, Radicioni2021b,mattei2022bowtie}, for the detection of non-trivial fluxes of disinformation on Twitter ~\cite{Caldarelli2021,DeClerck2022a,DeClerck2022b}, for highlighting coordinated behaviours among Twitter automated accounts~\cite{caldarelli2020role, Bruno2022}, for the identification of different information diets on Facebook~\cite{Guarino2021} or for the detection of the effective subjects of debate~\cite{Radicioni2021a, Radicioni2021b,mattei2022bowtie}. 
The main idea is to create a maximally random benchmark (i.e., maximising the Shannon entropy associated with the system under analysis) that preserves some (topological) property of the original system. In this sense, with the aim of detecting non-trivial behaviours, maximum-entropy null-models represent a tool that, at the same time, is general and tailored on the observed network. 

% \subsection{Entropy-based null-models for complex network analysis}
% Here, we provide the sketch of the definition of the entropy-based null-models and  their implementation for detecting discursive communities on the retweet network. For more details on the entropy-based techniques for complex networks, we forward the interested reader to the review in~\cite{Cimini2018a} and to the references therein.

%\subsubsection*{Entropy-based null-models}
Here, we provide the sketch of the definition of the entropy-based null models for complex network analysis and all references for further information.
%the main flows of traffic i.e., those produced by the main communities of users on the retweet network. 

%\textbf{utenti verificati/non verificati che interagiscono tra loro in modo non bnale (l'interazione non banale porta alla clusterizzazione)}

%For more details on the entropy-based techniques for complex networks, we forward the interested reader to the review in~\cite{Cimini2018a} and to the references therein.

The aim of the entropy-based null-models is to define a benchmark for the analysis of a real network $G^*$ that 
is maximally random, but for a set of topological constraints $\vec{C}$ observed on $G^*$.
Thus, we define an \emph{ensemble} of graphs $\mathcal{G}$, i.e., the set of all possible graph configurations, from the empty graph to the fully connected one, all having the same number of nodes as in the real network. Then, we can assign a probability to every representative of the ensemble by maximising the relative Shannon entropy, i.e.,
\begin{equation*}
    S=-\sum_{G\in\mathcal{G}}P(G)\ln P(G),
\end{equation*}
under the constraint that the average over the ensemble of the vector $\vec{C}$ is exactly the value observed in the real network $G^*$, i.e., $\langle \vec{C}\rangle_\mathcal{G}=\vec{C}(G^*)$.
The result of this procedure returns in an Exponential Random Graph, i.e. $P(G)\sim e^{-\vec{C}(G)\cdot\vec{\theta}}$, where $\vec{\theta}$ are the Lagrangian multipliers associated to the constrained maximisation~\cite{Jaynes1957,park2004statistical}. The maximisation of the likelihood, i.e., the probability of observing the real system, is then implemented to find the numerical values of $\vec{\theta}$~\cite{Garlaschelli2008,Squartini2011a}.

Recently, a fast and efficient Python module able to solve many of the entropy-based null-models present in the literature was released and is available at \url{https://pypi.org/project/NEMtropy/}.%{\color{blue}\underline{Pypi}}.  %\newhref{https://pypi.org/project/NEMtropy/}{Pypi}~\cite{Vallarano2021}.

The importance of using a properly defined unbiased benchmark for the analysis of the spread of online disinformation was stressed in a recent work by De Clerck et al.~\cite{de2022maximum}:
the authors  show how different entropy-based null-models can highlight different features of the various disinformation campaigns. In this paper, we will consider the entropy-based null-model known as Bipartite Configuration Model
(BiCM~\cite{Cimini2018a,Saracco2015}) as a benchmark to maintain only verified Twitter accounts that have statistically significant interactions with unverified ones.
%,  thus focusing on the actual information exchanged in the network and getting rid of random noise. 
In Section~\ref{sssec:DisCo} we describe the use of this model as a component of our filtering procedure.
\section{Methods}

%sezione portata qua da Appendice
\subsection{Bipartite Configuration Model, Validated Projection and Community Detection}\label{sssec:DisCo}

%We apply the BiCM validation procedure~\cite{Cimini2018a,Saracco2015} to verified Twitter users. 

\begin{figure*}[h!]
  \caption{Entropy-based filtering procedure
  \label{fig:figure_1}}
  %\Description[XXX]{XXX}
  \includegraphics[scale=0.25]{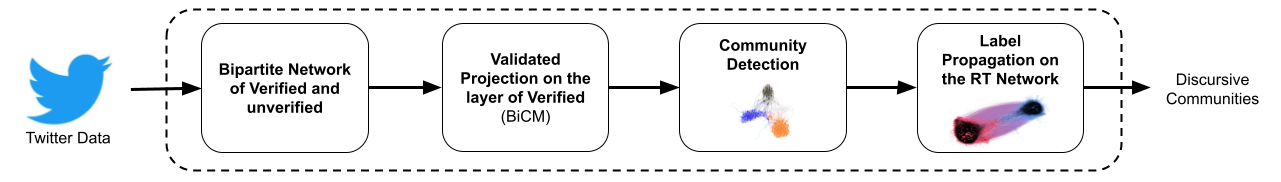}
\end{figure*}

Here, we describe how we filter accounts in our dataset using the validation procedure known in the literature as Bipartite Configuration Model BiCM~\cite{Cimini2018a,Saracco2015}.
As anticipated, our aim is to bring out political communities, leveraging the knowledge of the  political affiliation of verified users. The procedure that starts from raw Twitter data to obtain the discursive communities is depicted in Figure~\ref{fig:figure_1}.

The first observation is that most of the online debate is led by verified users, i.e., accounts whose owners are certified by the platform itself~\cite{Becatti2019d,Caldarelli2021,caldarelli2020role}. It is possible, therefore, to leverage this information to obtain proper communities of `similar' verified users: the intuition is that verified users with similar opinions in an online debate should have the same audience of `standard' users. 

Therefore we represent the retweet interactions between verified and unverified users as a bipartite network, i.e., networks in which nodes are divided in two sets, $\top$ and $\bot$ -called \emph{layers}- and connections are allowed only between layers; verified and unverified users are then represented by the two layers.

% As anticipated, the conceptual procedure of the entropy-based null-models can be applied for the practical detection of communities of users interacting among themselves via retweets, or \emph{discursive communities}~\cite{Becatti2019d,caldarelli2020role,Radicioni2021a}. \textbf{In particular we can leaverage on the concept of 'similarity' between verified users to extract community (that can be characterized from verified) that can be used to label others connected unverified users (la cui natura o similarita con altri utenti sarebbe sconosciuta). The basic idea is then to focus our analysis on the traffic flows produced by the main user communities \todo{spiegare communities basate su similarita'} participating in the discussion. Here, we describe our procedure for extracting these message flows.

% \man{The first observation is that most of the online debate is led by verified users,  i.e., accounts whose owners are certified by the platform itself. It is possible, therefore, to leverage this information to obtain proper discursive communities~\cite{Becatti2019d}: the intuition is that verified users with similar opinions in an online debate should have the same audience of `standard' users. Therefore we represent the retweet interactions between verified and unverified users as a bipartite network, i.e., networks in which nodes are divided in two sets, $\top$ and $\bot$ -called \emph{layers}- and connections are allowed only between layers; verified and unverified users are then represented by the two layers.}

We then project the bipartite network on the layer of verified users. Nevertheless, the projection only does not tell us so much: In fact, the common retweeters of two verified users could be many due to popularity of the latter or because the retweeters are retweeting many verified users.  We, therefore, need a benchmark that is maximally random and able to discount the effect of these two ingredients, which, in terms of the bipartite network defined above, are translated into the degree sequence of both layers. The entropy-based null-model for bipartite network discounting the information of the degree sequence is known as \emph{Bipartite Configuration Model} (BiCM,~\cite{Saracco2015}).

Using the BiCM as a benchmark, it is possible to validate the projection of the bipartite network on one of its layers: the co-occurrences observed in the real system are compared with the related BiCM distributions and, if they are statistically significant, they are validated~\cite{Saracco2017}. Therefore, the result of the validation procedure is a monopartite undirected unweighted network of verified users, in which two nodes are connected if the number of common retweeters is statistically significant, i.e., {\it it cannot be explained simply by the bipartite degree sequence.}

We subsequently run the Louvain community detection algorithm~\cite{Blondel2008} on the validated network of verified users to obtain the main communities. Each of these communities was manually labeled based on the characteristics of the  verified users inside.

%We subsequently run the Louvain community detection algorithm~\cite{Blondel2008} on the validated network of verified users to obtain the main communities. 
%\hlworkshop{Of all the communities of verified accounts that emerge, two have strictly political connotations, one Republican, the other Democratic.} 
%\hl{Two main tipology of communities emerge from the analysis. The biggest one is mostly composed by Republican supporters. The second most populated community is a mixed one, including Republicans, Democrats and some journals and journalists with various political orientations}.

%We subsequently run the Louvain community detection algorithm~\cite{Blondel2008} on the validated network: the so-obtained labels are then propagated on the retweet network using the Raghavanan et al. algorithm~\cite{Raghavan2007b}, in order to provide all users a discursive community label. Several works, like~\cite{Becatti2019d,caldarelli2020role,Radicioni2021a,DeClerck2022a,DeClerck2022b} show that the procedure above is particularly effective in capturing the structure of Twitter online debate.

Then, to include also unverified users, the so-obtained labels are propagated on the retweet network using the Raghavanan et al. algorithm~\cite{Raghavan2007b}, in order to provide all users a %discursive 
community label. Several works, like~\cite{Becatti2019d,caldarelli2020role,radicioni2021networked,DeClerck2022a,de2022maximum}  
show that the procedure above is particularly effective in capturing the structure of Twitter online debate.
The communities identified using this procedure are termed \textit{discursive communities}. These communities comprise both verified and non-verified users who actively contribute to the development of a shared discourse. A pivotal aspect of our characterization process begins with the verified users within these communities. This approach allows us to effectively ascertain whether a discursive community possesses a political nature and adheres to a specific orientation.

\subsection{Article's reliability measure through NewsGuard}
%All the domains in our Twitter dataset have been tagged according to their degree of credibility and transparency,  as indicated by the browser extension and mobile app NewsGuard\footnote{\url{https://www.newsguardtech.com/}}. The NewsGuard initiative was born from the joint effort of journalists and software developers, aiming at evaluating news sites according to  criteria concerning credibility and transparency. For evaluating the credibility level, the metrics consider, e.g., whether the news source regularly publishes false news, does not distinguish between facts and opinions, does not correct a wrongly reported news. For transparency, instead, the toolkit takes into account, e.g., whether owners, founders or authors of the news source are publicly known, and whether advertisements are easily recognizable\footnote{Details on the procedure for the  evaluation  are available at: \url{https://www.newsguardtech.com/ratings/rating-process-criteria/}.}

% Here, we report a series of analyses related to the domains that mostly appear in the tweets of the validated network of verified users. 

In this study, our evaluation of untrustworthy content takes a distinct approach by centering the attention on the source of the content, the publisher, rather than focusing solely on individual stories. We categorize each source based on its credibility and transparency, a process that relies on the assessment conducted by NewsGuard\footnote{\url{https://www.newsguardtech.com/}}. This assessment involves expert journalists annotating online news sources.

This source-centric approach is grounded in the belief that the intent and practices of the publisher play pivotal roles in determining the trustworthiness of news~\cite{lazer2018science}. Furthermore, examining every single article individually is impractical, which makes this approach highly favored as it facilitates large-scale studies~\cite{Caldarelli2021, shao2018spread}.

% In this study, we assess the presence of untrustworthy content by directing our focus towards the original source, the publisher, rather than individual stories. Each source is categorized based on its credibility and transparency, using the evaluation conducted by NewsGuard\footnote{\url{https://www.newsguardtech.com/}}, which involves expert journalists annotating online news sources. This source-based approach is based on the notion that the intent and processes of the publisher are the key factors in determining untrustworthy news~\cite{lazer2018science}. Additionally, evaluating every single article is not feasible, making this approach widely favoured for enabling large-scale studies~\cite{Caldarelli2021, shao2018spread}.
% } 
%The NewsGuard initiative was born from the joint effort of journalists and software developers, aiming at evaluating news sites according to criteria concerning credibility and transparency.

For evaluating the credibility level, the Newsguard metrics consider, e.g., whether the news source regularly publishes false news, does not distinguish between facts and opinions, does not correct a wrongly reported news. For transparency, instead, the toolkit takes into account, e.g., whether owners, founders or authors of the news source are publicly known, and whether advertisements are easily recognizable
\footnote{Details on the procedure for the  evaluation  are available at: \url{https://www.newsguardtech.com/ratings/rating-process-criteria/}}.

% As a first step, we considered the network of verified accounts, whose communities and subcommunities have been shown
% in Figure~\ref{fig:subcomm_netwk}. On this topology, we labelled all domains that had been shared at least 20 times in tweets and retweets. 
\begin{table}[ht!]
\caption{Tags for domain reputation labeling.\label{table:domains-tags}}
%OLD
%\caption{Tags for domain reputation labeling. Tags are inherited from NewsGuard, the UNC tag indicates that NewsGuard has not yet tagged that domain. \label{table:domains-tags}}
\centering
\begin{tabular}{c|l}
label & \text{description}\\
\hline
\hline
T & Trustworthy news domain\\
%$\sim\text{R}$ & Quasi Reputable news source\\
N & untrustworthy news domain\\
P & Platform (e.g., reddit.com, twitter.com)\\
S & Satire\\
UNC & unclassified\\
\hline
\end{tabular}

\smallskip

\end{table}

Table~\ref{table:domains-tags} shows the tags that NewgsGuard associates with each news domain. Since we are interested in quantifying the reputation of news domains publishing during the period of interest,  
 we do not consider sources corresponding to platforms (tag P). We will also not consider satirical news (tag S). 
The tags T and N in Table~\ref{table:domains-tags} are used only for news sites, be they newspapers, magazines, TV or radio channels, and stand for trustworthy and non-trustworthy, respectively. 
We clarify that for us, a domain corresponds to the so-called `second-level domain'\footnote{\url{https://en.wikipedia.org/wiki/Domain\_name}}, i.e. the name directly to the left of .com, .net and all other top-level domains. For example, \url{nytimes.com} and \url{latimes.com} are considered domains in this manuscript.

\subsection{Bot detection}
\label{sec:botdetection}
The accounts in our dataset were examined using the bot detector Botometer, \man{one of the most well-known bot detection tools in the literature}~\cite{Varol2017,FerraraArming2019,DBLP:conf/cikm/Sayyadiharikandeh20,yang2020scalable}. The tool is based on a supervised machine learning approach using Random Forest classifiers~\cite{Breiman2001}. 
%(in its earliest version known as `BotOrNot', publicly available as web interface and API since May, 2014) 
We  rely on Botometer v4, the new version of the bot detector, which has been shown to perform well for
detecting both single-acting bots and coordinated campaigns~\cite{DBLP:conf/cikm/Sayyadiharikandeh20,yang2020scalable}. 
In particular, we have adopted Botomoter v4 premium in the lite version BotometerLite\footnote{\url{https://cnets.indiana.edu/blog/2020/09/01/botometer-v4/}}, which does not interface with Twitter, but simply takes the tweet, retrieves the author, and does the necessary follow-up analysis. This light version only needs the information in the user profile to perform bot detection, so it can also process historical data published by accounts that are no longer active. Each request to BotometerLite can process a maximum of 100 users, with a limit of 200 requests per day, resulting in a maximum of 20k account checks per day. 
%(50 dollars/month)
% allows a rate limit of 17,280 requests per day (each request processes 1 user only).  
%  so this allows to check more than 1,700,000 users per day). 
% In addition, v4 premium offers the lite version BotometerLite, which does not interface with Twitter, but simply takes the tweet, retrieves the author, and does the necessary follow-up analysis. 
The immediate output of Botometer is the bot score $S$, which ranges over \{0, \ldots 1\}, but does not represent the probability that the considered account is a bot. The value needs to be compared with other scores within a group of accounts to come up with a plausible ranking.

%In the main text,  we consider the distribution of bot scores on the main discursive communities emerged from data. 

\section{Results}

\subsection{Dataset}\label{sec:Dataset}
%The dataset can be downloaded at \url{https://doi.org/10.7910/DVN/ANBPTC}, hosted at \url{https://dataverse.harvard.edu/dataset.xhtml?persistentId=doi:10.7910/DVN/CUWN54}.

Using the Streaming Twitter API, we collected around 5.3M tweets in the week immediately preceding the elections (27 October-3 November 2020). To guide the data collection, we chose keywords 
%(see Appendix, Table~\ref{tab:keywords})
combining the name of four swing and four safe states (see Table~\ref{tab:dataset_by_states}) with the candidates (i.e., Trump and Biden).

\begin{table}[ht!]
\caption{Keywords which drove the data collection phase. \label{tab:keywords} }

\begin{tabular}{l}
\hline
Keywords\\
\hline
arizona biden\\
arizona trump\\
florida biden\\
florida trump\\
michigan biden\\
michigan trump\\
pennsylvania biden\\
pennsylvania trump\\
new jersey biden\\
new jersey trump\\
indiana biden\\
indiana trump\\
washington biden\\
washington trump\\
louisiana biden\\
louisiana trump\\
\hline
\end{tabular}
\end{table}

%\mari{The choice of the eight states was made taking into account polls and opinions of pundits and experts in the field in the months before the 2020 election\footnote{\url{https://www.maynoothuniversity.ie/research/spotlight-research/10-swing-states-will-decide-us-presidential-election}}. As for the states that were declared safe by these studies, we considered two traditionally associated with Democrats (Washington and New Jersey) and two with Republicans (Indiana and Louisiana). For the states that those polls gave as swing, we considered as a selective factor the number of inhabitants}. 

The states were selected based on measures and indications provided in reports by experienced political analysts in the months leading up to the 2020 elections\footnote{\url{https://www.cookpolitical.com/analysis/national/national-politics/latest-cook-political-report-electoral-college-map}\label{foot:cookpolitical}}. We chose a balanced list of states, four safe states and four swing states.
For the safe states, we chose two pairs that were balanced in terms of political leanings and presidential electors. We took Washington and New Jersey from the solid Democratic states and Indiana and Louisiana from the solid Republican states. This results in 26 electoral votes for the Democratic candidate and 19 for the Republican. For the selection of the swing states, we took the three most important states from the point of view of presidential electors: Florida (29 votes), Pennsylvania (20 votes), and Michigan (16 votes); we also added Arizona (11 votes) because it has been of particular interest in the election debates\footnote{\url{https://fivethirtyeight.com/features/how-arizona-became-a-swing-state/}}$^,$\footnote{\url{https://www.washingtonpost.com/politics/2022/09/16/senate-control-midterm-elections-2022/}}.
\man{We should further clarify that our choice was not driven by any ``formal'' definition based on statistics relative to the results of previous elections, but by the indications of political analysts (especially those in \footref{foot:cookpolitical}). For example, Arizona gave its electoral votes to the Republicans in the 2000, 2004, 2008, 2012, and 2016 elections. However, the consensus among political analysts, based on various polls, was that Arizona was no longer a safe state for Republicans. In fact, Arizona gave its electoral vote to Biden in 2020. With that in mind, we considered Arizona a swing state, even though the historical data would have placed it in the safe set.}

%The complete list of pairs of keywords sought for data collection is in the Appendix,  Table~\ref{tab:keywords}. 

The data was further processed to (i)
identify user communities with a vested interest in the political narrative through our filtering process, (ii) classify link domains using NewsGuard, and (iii) map each tweet to its corresponding state type (i.e., swing or safe) using a content-based approach.

% identify (through our filtering procedure) the communities of users interested in the political narrative,} 
% (ii) enable link domain classification through NewsGuard, and (iii)  find a mapping between each tweet and the kind of state (i.e., swing or safe). % and (iv)  filter out useless data from the entire set.  

The procedure for filtering the data set is described in Section~\ref{sssec:DisCo}. From here on, we will refer to the product of the filtering procedure as the `validated dataset' (to distinguish it from the original dataset).
For both the verified and unverified accounts that pass the filtering procedure, we also collect the bot scores via BotometerLite.

For URL classification, we rely on NewsGuard\footnote{\url{https://www.newsguardtech.com/}}, which provides a set of \{\emph{domain\_name}, \emph{tag}\} pairs (tags are in Table~\ref{table:domains-tags}). It was therefore necessary to translate all the short-form URLs contained in the text of the tweets, so that we could have the domain names in clear. 
%We clarify that a domain, for us, corresponds to the so-called `second-level domain' name\footnote{\url{https://en.wikipedia.org/wiki/Domain_name}}, i.e., the name directly to the left of .com, .net, and any other top-level domains. For instance,  \url{nytimes.com} and \url{latimes.com} are considered as domains in the present manuscript.

We use a content-based approach to establish the association between each tweet and the state type (i.e., swing or safe). In practice, we first check each tweet - or retweet - for the presence of at least one state name from the selected list (e.g., Arizona, Florida, etc.). We then exclude any tweets that contain more than one state name (approximately 1.5 million tweets contain more than one state name). Consequently, each tweet in the resulting dataset contains only one state name, which can be swing or safe. \man{Although it is true that posts discussing the states under analysis have been lost, we prefer such a conservative approach to eliminate possible noise from our data set.}
Furthermore, we only consider English tweets (non-English tweets number about 422,000). The resulting dataset contains about 3.3 million tweets and about 398,000 URLs (see Table~\ref{tab:dataset_by_states}).

\man{Finally, we conclude this subsection by noting that our analysis aims to investigate whether the debate about swing states is meaningfully different from the debate about safe states. Therefore, we are not interested in checking the origin of the tweets, i.e. whether or not their authors are located in the states under analysis. In fact, even users outside the US can contribute to the debate and/or the level of disinformation in it.}
    
% We employ a keyword-based approach to find the association between each tweet and the state type (i.e., swing or safe). In practice, from the text of each tweet -or retweet-, we first check for the presence of at least one state name among the chosen ones (Arizona, Florida, etc.) and then we discard all the tweets in which more than one state appear \man{(the tweets with more than one state name were $\sim1.5$M)}. Each tweet in the resulting dataset, thus, contains only one state name, which can be either swing or safe. In addition, we consider English tweets only \man{(not-English tweets were $\sim422$k)}. The resulting dataset consists of $\sim$3.3M tweets and $\sim$398k URLs (see Table~\ref{tab:dataset_by_states}).

\begin{table}
\caption{Twitter's statistics by state. The asterisk `$\ast$' indicates swing states.}
\begin{tabular}{lcc}
\hline
State &  No. Tweets &  No. URL \\
\hline

Arizona$\ast$ &      224046 &    34637 \\
Florida$\ast$  &      744006 &    85373 \\
Michigan$\ast$ &      734600 &    87529 \\
Pennsylvania$\ast$ &     1209083 &   145067 \\

New Jersey &       38007 &     8114 \\
Indiana &       17185 &      988 \\
Washington &      342104 &    36254 \\
Louisiana &        6886 &      633 \\
\hline
Total & 3315917 & 398595 \\
\hline
\end{tabular}
\label{tab:dataset_by_states}
\end{table}

%\subsection{RQ1 (presence of political community)}
\subsection{Detection of discursive communities}
\label{sec:politicalcomm}

\begin{table*}[ht!]
\caption{Characteristics of the main discursive communities. \label{tab:community_characterization}}
%OLD
%\caption{Characteristics of the main discursive communities. 2 main communities emerge, {\sc Rep} and {\sc Rep-Dem-Journ}. With {\sc Others} we characterize all the accounts that do not belong to the giant component: their contribution is going to be disregarded in the following, since they do not contribute  to the entire debate.\label{tab:community_characterization}}
\resizebox{\textwidth}{!}{\begin{tabular}{lccccccc}
\hline
           Community &  No. Users &  No. Tweets &  Tweets Safe &  Tweets Swing &  No. URL &  Left &  Right \\
 \hline
 {\sc Rep} &     269019 &     2083158 &           12.35 &            87.65 &   241488 &     0.59 &     \textbf{49.74} \\
 {\sc Rep-Dem-Journ} &     213679 &      919949 &           10.39 &            89.61 &    92412 &    \textbf{16.18} &      1.86 \\

 {\sc Journ-1} &        197 &        1174 &            4.86 &            95.14 &      485 &     3.30 &      0.82 \\
 {\sc Journ-2} &         53 &         404 &           10.64 &            89.36 &       74 &    22.97 &      2.70 \\
 {\sc Others} &     218880 &      311232 &           16.44 &            83.56 &    64136 &     6.62 &     14.60 \\
 \hline
 Dataset &     701828 &     3315917 &           12.19 &            87.81 &   398595 &     5.18 &     32.92 \\
 \hline
 \end{tabular}}
 \end{table*}

We execute the procedure outlined in Section~\ref{sssec:DisCo} to identify discursive communities, which are groups of Twitter accounts that actively contribute to the development of a shared discourse by retweeting among themselves;
results are summarised in Table~\ref{tab:community_characterization}. 

The giant component of the retweet network includes more than $4.8\times10^{5}$ accounts, while nearly $2.2\times10^{5}$ accounts belong to smaller clusters: the latter are not going to be analysed in the following since they are not relevant for the entire debate. 

To characterize the community structure of the giant component, we conduct a manual analysis \emph{a posteriori} of the various communities, leveraging the presence of verified accounts (as discussed in Section~\ref{sssec:DisCo}) i.e., authentic public interest accounts like politicians, journalists or VIPs.

To assign labels to the list of verified users within each community, we gave priority to users with higher node degrees, indicating a greater number of connections. 
The largest community within the giant component primarily consists of Republican supporters and comprises approximately $2.7\times10^{5}$ users. Some examples of users within this community include `@TrumpWarRoom', `@TeamTrump', and `@TrumpStudents'. This community will be referred to as {\sc Rep} henceforth.

The second most populated community, with around $2.1\times10^{5}$ accounts, is a mixed one encompassing Republicans, Democrats, as well as various journals and journalists with diverse political orientations. Accordingly, it will be labeled as {\sc Rep-Dem-Journ}. 

Accounts in the {\sc Rep} and {\sc Rep-Dem-Journ} communities are responsible for over 90\% of the tweets in our dataset (\man{see Table~\ref{tab:community_characterization}}). While other communities do exist within the giant component, their size is practically negligible compared to the ones described above or they lack a clear political orientation. Therefore, they will not be considered in the subsequent analysis.

\begin{figure*}[h!]
  \caption{Retweet Network after label propagation (547k nodes, 1.8M edges).\label{fig:retweet_network}}
  %OLD
  %\caption{Retweet Network after label propagation (547k nodes, 1.8M edges). The two main communities that emerge are the red one, characterized by Republican supporters, and the blue one, a mix of Republicans, Democrats, and journalists of various affiliations.\label{fig:retweet_network}}
  
  %\Description[Retweet Network after label propagation]{After label propagation on the retweet network, two main communities that are focused on political narrative emerge}
  \includegraphics[scale=0.2]{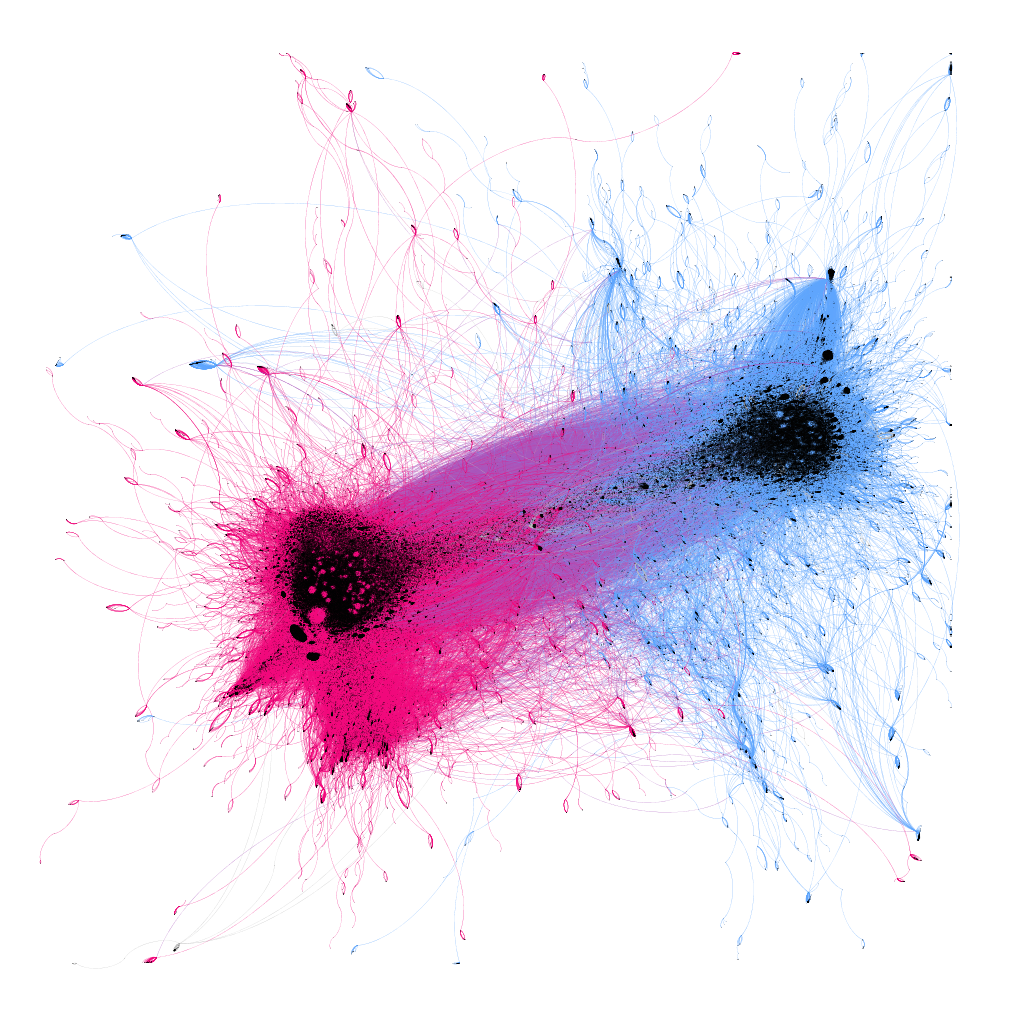}
\end{figure*}

In particular, our analysis will focus only on the result of the entropy-based filtering procedure, i.e. the users belonging to the {\sc Rep} and {\sc Rep-Dem-Journ} communities. \man{Figure~\ref{fig:retweet_network} shows these two main communities, which emerge after running the label propagation algorithm in \cite{Raghavan2007b} to the retweet network}.

% To characterize the community structure of the giant component, we manually analyse \emph{a posteriori} the various communities, leveraging the presence of verified accounts (see Section~\ref{sssec:DisCo}). \man{To assign labels to the list of verified users within each community, we prioritize users with higher node degrees, indicating a greater number of connections.} The most crowded community of the giant component is mostly composed by Republican supporters and it includes $2.7\times10^{5}$ users. \man{Some examples of users within this community include `{\it TrumpWarRoom}', `{\it TeamTrump}', and `{\it TrumpStudents}'}. In the following, this community is going to be called {\sc Rep}. The second most populated community ($\sim2.1\times10^{5}$ accounts) is a mixed one, including Republicans, Democrats and some journals and journalists with various political orientations and, therefore, it will be called {\sc Rep-Dem-Journ}. Accounts in {\sc Rep} and {\sc Rep-Dem-Journ} communities are responsible of more than 90\% of the tweets of our dataset (Figure~\ref{fig:retweet_network} shows the resulting retweet network). Other communities are present in the giant component, but, since their dimension is practically negligible with respect to the ones just described \man{or have not a clear political orientation}, they are not going to be considered in the following. \man{In particular, from here in our analysis we only consider the product of the entropy-based filtering procedure i.e., the users belonging to the {\sc Rep} and the {\sc Rep-Dem-Journ} communities. 
% }

The result of the analysis of news domains in the tweets of major communities is a good indication of their correct labeling. The Left and Right columns in Table~\ref{tab:community_characterization} represent the percentage of sources identified by NewsGuard as left-wing and right-wing oriented, respectively.
Within the {\sc Rep} community, almost 50\% of the shared URLs come from right-leaning sources. In the {\sc Rep-Dem-Journ} community, the prevalence of left-leaning sources is significantly lower, approximately 16.2\%. This measure can probably be explained by observing the mixed composition of the {\sc Rep-Dem-Journ} community.%Due to the mixed composition of the {\sc Rep-Dem-Journ} community, the prevalence of left-oriented sources is notably lower, at approximately 16.2\%.

\man{We also conducted a manual analysis of a sample of users from the emerging communities to verify the correct composition of the latter.
Specifically, we randomly selected a subset of 99 unverified accounts and created a balanced sample representative of emerging discursive communities, including Rep, Rep-Dem-Journ, and those users assigned to other communities or without any community association. We then manually annotated the Twitter users within the sample, taking into account (i) the content of the messages they write or retweet, (ii) the political orientation of the news publishers they share (using the NewsGuard labels), and (iii) the political orientation of well-known Twitter users they retweet.

When comparing the manual annotations with our labeling procedure, 89 out of 99 users showed consistent labels. However, 10 users who were not assigned to any community by our filtering procedure showed inconsistencies in the labels. A content analysis of these 10 users reveals an association with Republican and Democratic political visions (6 and 4 users, respectively). 
It is important to emphasize that these minor discrepancies do not affect our analysis, which focuses on users belonging to the main communities, i.e., {\sc Rep-Dem-Journ} and {\sc Rep}.}

\subsection{Reputation of news domains}\label{sec:reputationComm}

%\begin{table}[h!]
%\caption{Tags for domain reputation labeling. Tags are inherited from NewsGuard, the UNC tag indicates that NewsGuard has not yet tagged that domain. %(UNC = Unclassified)  
%\label{table:domains-tags}}
%\centering
%\begin{tabular}{c|l}
%\hline
%Label & Description\\
%\hline
%T & Trustworthy news domain\\
%$\sim\text{R}$ & Quasi Reputable news source\\
%N & Non-trustworthy news domain\\
%P & Platform (e.g., reddit.com, twitter.com)\\
%S & Satire\\
%UNC & unclassified\\
%\hline
%\end{tabular}

%\smallskip

%\end{table}

%Table~\ref{table:domains-tags} shows the tags associated to the domains provided by NewsGuard. We are interested in quantifying reputation of news domains publishing during the period of interest. 
%Thus, we do not consider those sources corresponding to platforms (tag P). Also, we will not consider satiric news (tag S). 
% \fasa{; nevertheless, the information regarding their frequency are available for the interested readers in the Supplementary Material.} 
%Tags T and N in Table~\ref{table:domains-tags} are used only for news sites, be them newspapers, magazines, TV or radio social channels, and they stand for Trustworthy and Non-trustworthy, respectively. 

% Here, we report a series of analyses related to the domains that mostly appear in the tweets of the validated network of verified users. 
\begin{figure}[h]
  \centering
  \caption{Classification of links\label{fig:url_reliability_distribution}}
    %OLD
    %\caption{Classification of links: trustworthy news publisher (T); untrustworthy news publisher (N); not a news site, e.g. platforms as amazon.com (P); link corresponding to a domain not classified by NewsGuard (UNC).\label{fig:url_reliability_distribution}}
  %\Description[Classification of links according to NewsGuard rating]{The links belonging to the dataset and the validated set are classified according to NewsGuard rating}
  %\includegraphics[width=.47\textwidth]{./figure_3_left.png}\hfill
  %\includegraphics[width=.47\textwidth]{./figure_3_right.png}\hfill
  \includegraphics[width=.90\linewidth]{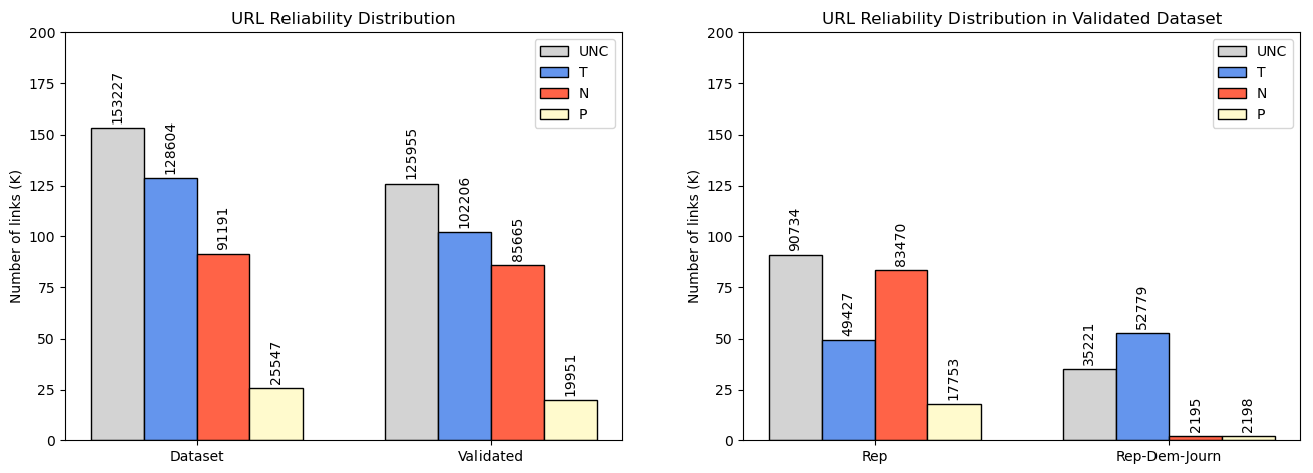}

\end{figure}

%\sout{We clarify that a domain, for us, corresponds to the so-called `second-level domain' name\footnote{\url{https://en.wikipedia.org/wiki/Domain_name}}, i.e., the name directly to the left of .com, .net, and any other top-level domains. For instance,  \url{nytimes.com} and \url{latimes.com} are considered as domains in the present manuscript.}

%Figure~\ref{fig:url_reliability_distribution} \hl{shows the distribution of URLs found in tweets in the complete and validated dataset regarding the tags that NewsGuard assigned to the related domains. A link is labeled UNC if the corresponding domain was not analyzed by NewsGuard, at least at the time of the data collection. T represent a trustworthy news publisher, N a non-trustworthy news publisher, and P is a link whose domain is not a news publisher.} \man{}

% commento present nella version workshop
Figure~\ref{fig:url_reliability_distribution} (left) shows the distribution of URLs found in tweets in the full and validated datasets, respectively. The URLs have been tagged according to the NewsGuard labels. The validation procedure discards $\sim17\%$ of the tweets with URLs from the full dataset, which is about 64k tweets. Thus, most of the links are distributed within the political communities that emerge from the data.

%A link is labeled UNC if the corresponding domain was not analyzed by NewsGuard, at least at the time of the data collection. T represent a trustworthy news publisher, N a non-trustworthy news publisher, and P is a link whose domain is not a news publisher.} \man{}

%\man{In particular, as seen in Section~\ref{sec:politicalcomm}, two main discursive communities emerge from the online debate, i.e., {\sc Rep} and {\sc Rep-Dem-Journ}. To understand whether there are differences within these two communities in terms of number and distribution of links to non-trustworthy news sites, we classify the news publishers inheriting NewsGuard tags for those sites (reported in the Appendix, see Table~\ref{table:domains-tags}).

%The {\sc Rep-Dem-Journ} community is, in terms of traffic, less numerous than {\sc Rep} (less than half a traffic in terms of tweets and much less than half in terms of URLs, see Table~\ref{tab:community_characterization}).

Figure~\ref{fig:url_reliability_distribution} (right) shows that, with respect to the entire dataset, 93\% of untrustworthy links (N) are shared within the two main political communities. In particular, about 91\% of the total is shared within the {\sc Rep} community. Furthermore, the links in the Rep community with publishers tagged as N by NewsGuard are mostly right-leaning (i.e., in terms of numbers we found Slightly Left $2$, Far Left $6$, Slightly Right $4831$, Far Right $76161$ links).

\begin{figure}[h!]
\centering
\caption{Distribution of the number of link sharing in {\sc Rep} (left) and {\sc Rep-Dem-Journ} (right) (see Table\ref{table:domains-tags}).\label{fig:sharing_by_reliability}}
%OLD
%\caption{Distribution of the number of link sharing in {\sc Rep} (left) and {\sc Rep-Dem-Journ} (right) for T, N and all Tags (see Table\ref{table:domains-tags}). Values are normalized applying a log function. The green line is the mean, the purple line is the median; outliers are not reported. \man{The box extends from the first quartile to the third quartile. The whiskers extend from the box to the outermost data point within 1.5 times the interquartile range from the box.}\label{fig:sharing_by_reliability}}
\includegraphics[width=.90\linewidth]{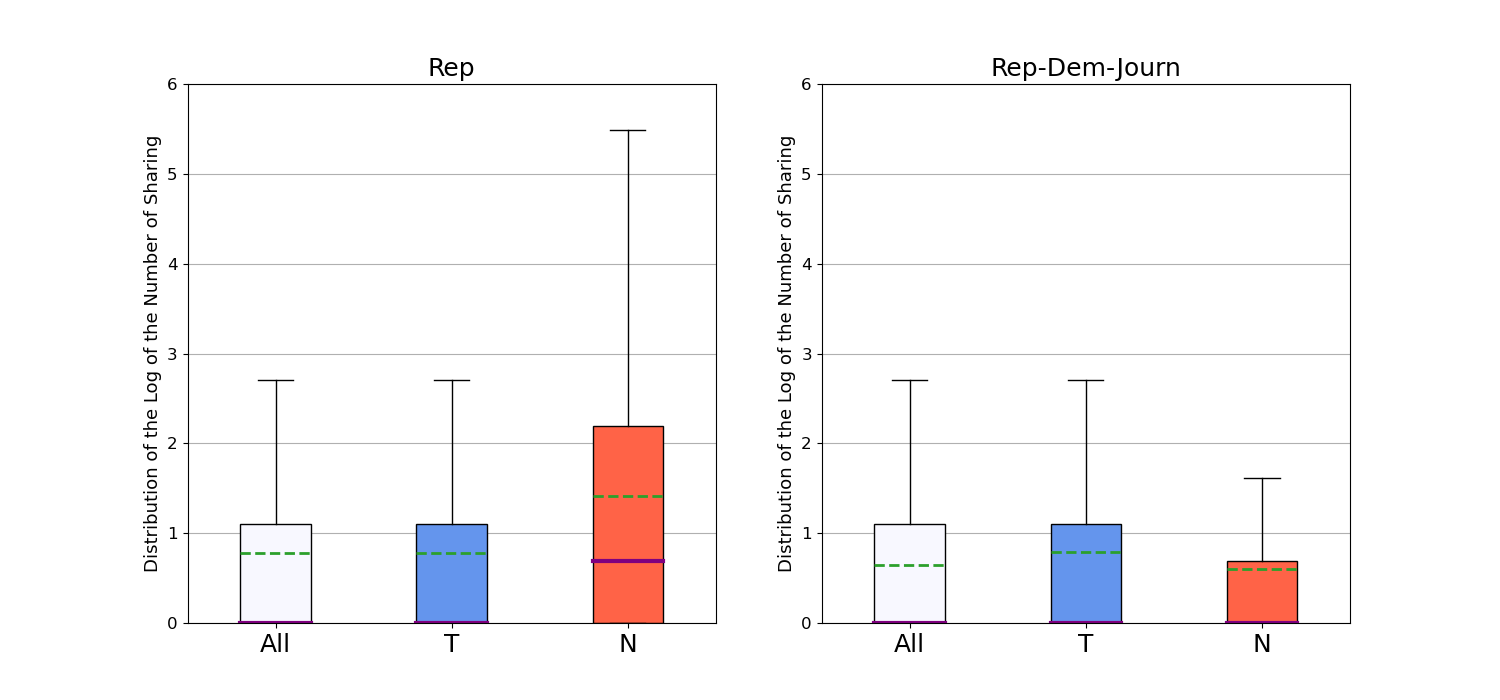}
\end{figure}

Figure~\ref{fig:sharing_by_reliability} shows the virality of the links, that is, how many times the links in our dataset have been shared. 
We can see that in {\sc Rep}, links of type N are shared many more times than other types of links. Specifically, in {\sc Rep}, an N link is shared on average 57 times, while in {\sc Rep-Dem-Journ} it is shared 7 times. These results suggest that untrustworthy links find fertile ground among Republican supporters.

\subsection{Reputation of news domains in tweets associated to swing and safe states}
\label{sec:reputationState}

\begin{table}[h!]
%commentate caption per versione da workshop
%\caption{Statistics regarding number of accounts, tweets, and type of URLs, \sout{per community and} per state type in \sc{Validated dataset}. 
\caption{Statistics show accounts, tweets, URL per state type in validated dataset, and discursive communities.
%Statistics regarding number of accounts, tweets, and type of URLs per state type in the validated dataset and in the two main discursive communities.
\label{tab:misinformation_in_swing_and_safe}}
\begin{tabular}{lccccc}
\hline
States &  No. Users &  No. Tweets &  No. URL &  T &  N \\%&  UNC \\
\hline

%\multicolumn{6}{l}{\sc{Dataset}} \\

%Swing &     636125 &     \textbf{2911735} &   352606 & \textbf{29.84} & \textbf{23.47} \\ %&   39.76 \\
%Safe &     214329 &      \textbf{404182} &    45989 & \textbf{50.87} & \textbf{18.33} \\ %&   28.37 \\

%commentate la riga sotto per versione da workshop
\multicolumn{6}{l}{\sc{Validated}} \\

Swing &     451840 &     \textbf{2649642} &   299210 & \textbf{28.30} & \textbf{26.06} \\ %&   39.76 \\
Safe &     170644 &      \textbf{352863} &    34586 & \textbf{50.66} & \textbf{22.25} \\ %&   28.37 \\

%commentate le righe sotto per versione da workshop
 \hline
 \multicolumn{6}{l}{\sc{Rep}} \\
 Swing &     251615 &     1825320 &   218565 & 18.66 & 34.72 \\ %&   38.76 \\
  Safe &     112556 &      257236 &    22819 & 37.92 & 33.25 \\ %&   26.40 \\
 
 \hline
 \multicolumn{6}{l}{\sc{Rep-Dem-Journ}} \\
 Swing &     200225 &      824322 &    80645 & 54.45 &  2.59 \\ %&   40.45 \\
 Safe &      58088 &       95627 &    11767 & 75.37 &  0.91 \\ %&   22.11 \\
\hline
\end{tabular}
\end{table}
Here, we analyze the flow of disinformation in tweets associated with swing or safe states and per discursive community. We recall that a tweet is associated with a state if the name of the state is present in the tweet text. By construction, each tweet in our dataset contains only one state name.

% In questa sezione analizziamo il dataset dividendo il traffico in base alla tipologia di stato i.e., swing or safe. Un tweet viene associato alla categoria swing se nel testo del tweet (o del retweeted) e presente il nome di uno stato swing (vedi Tab \ref{tab:dataset_by_states}); analogamente viene fatto per i safe. Ricordiamo che per costruzione, ciascun tweet del nostro dataset puo contiene un solo nome di stato (che puo appartenere alla categoria swing or safe).

% In modo simile a come fatto in Section \ref{sec:reputationComm}, le nostre analisi si focalizzeranno dapprima sul capire come il traffico un-reputable (and reputable) si suddivide (i) tra swing e safe states e (ii) tra swing e safe considerando anche la community politica di appartenenza del tweet (i.e., {\sc Rep} or {\sc Rep-Dem-Journ}); cerceremo poi di evidenziare le eventuali differenze in termini viralita nella condivisione.

Table~\ref{tab:misinformation_in_swing_and_safe} gives statistics on the number of accounts, tweets, and URLs related to the kind of state associated with the tweets and to the two main political communities.
We see that the vast majority of traffic is associated with tweets about swing states (about 88\% of the total, see row {\sc Validated}, column No. Tweets). When looking at links pointing to untrustworthy news sites (N), the concentration for swing states - 26.06\%.
 - is higher than for safe states - 22.25\%%18.33\% !!! sbagliato
. The  concentration of trustworthy links (T) is higher for safe states - 50.66\% % !!! sbagliato 50.87\%
{\it vs} 28.30\% % !!! sbagliato 29.84\%
for swing states.

To statistically validate the frequencies of N and T links in tweets associated with swing and safe states, we performed the chi-square~\cite{Pearson1900} statistical test. The comparison between the frequency distribution of T and N links in such tweets and the frequency distribution of T and N links in the validated dataset is significantly different: the obtained p-value for the test is below the order of $10^{-65}$.

% DARIVEDERE
% We applied the chi-square test to assess the statistical significance of our results. The sharing frequencies of links N (as well as links T and UNC) were found to be significantly different from the frequencies observed in the {\sc Validated} dataset for tweets associated with the categories `swing' and `safe'. Furthermore, we observed that the sharing frequencies of links T, N, and UNC associated with the categories `safe' and `swing' are significantly different from the expected frequencies considering the entire dataset. The obtained p-values for these tests were all below the order of $10^{-65}$.

% commentato per versione da workshop
At the community level, in agreement with the results in Section~\ref{sec:reputationComm}, we observe a higher concentration of links N in the {\sc Rep} community; however, we do not observe substantial differences in terms of percentage of links N between swing and safe states for both {\sc Rep} and {\sc Rep-Dem-Journ} (Table~\ref{tab:misinformation_in_swing_and_safe}, rightmost column).  For both communities, the highest concentration of trustworthy links (T) is in tweets associated with safe states (column T).

\begin{figure}[h]
  \centering
   \caption{Distribution of the number of link shared per kind of state in {\sc Rep}.}\label{fig:url_sharing_behaviour_in_swing_and_comm}
   %OLD\caption{Distribution of the number of link shared per kind of state in {\sc Rep}. Values are normalized applying a log function. The green line is the mean, the purple line is the median.\man{The box extends from the first quartile to the third quartile. The whiskers extend from the box to the outermost data point within 1.5 times the interquartile range from the box.}}\label{fig:url_sharing_behaviour_in_swing_and_comm}
   \includegraphics[width=.90\linewidth]{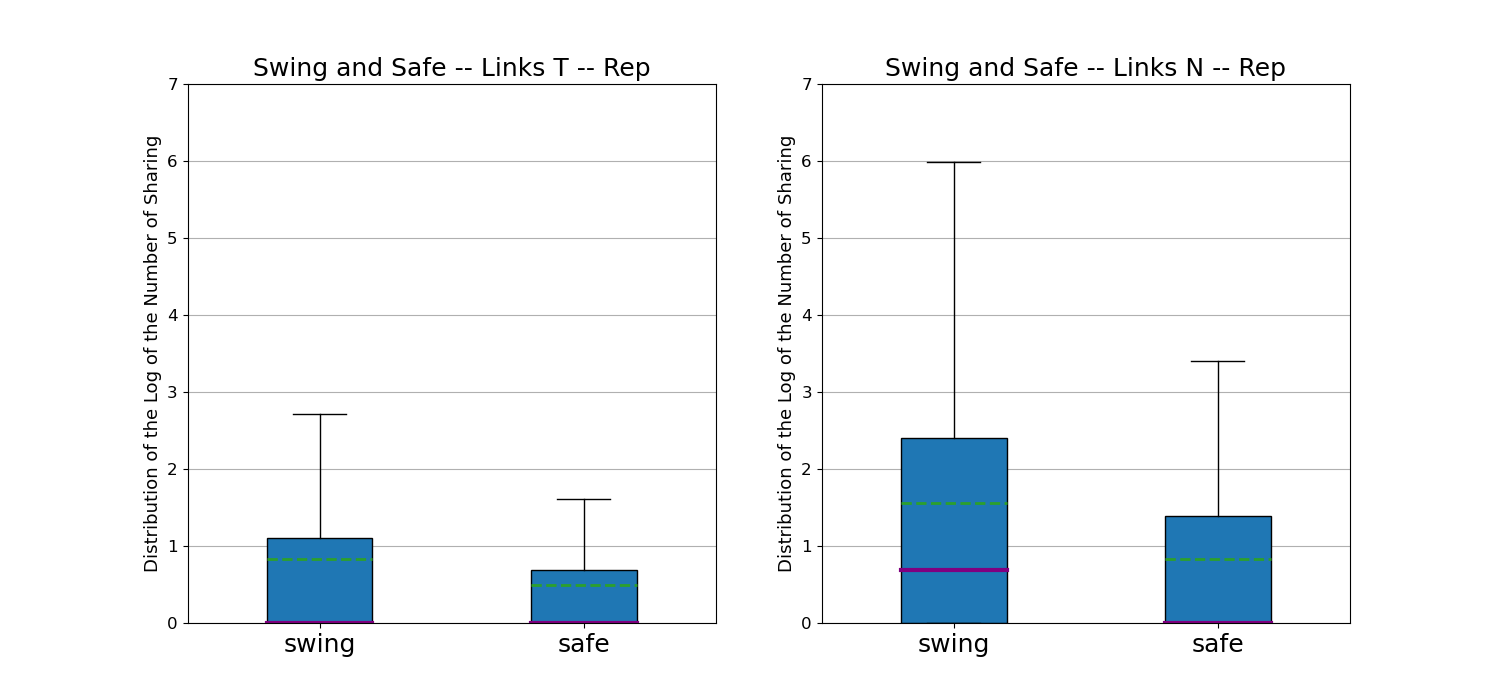}
 \end{figure}

Figure~\ref{fig:url_sharing_behaviour_in_swing_and_comm} shows that in {\sc Rep}, untrustworthy links (N) are shared many more times on average in the debate associated with swing states. Specifically, the average number of shares is 66 times for \emph{swing} and 22 times for \emph{safe}. For trustworthy links (T), a similar but not as pronounced behavior is observed.

%$[$ TODO or not $]$ Focus on swing state - N links - rep community, Figure\ref{fig:sharing_behaviours_swing_rep}: TODO potrebbe essere carina un analisi in piu.

\subsection{Social bots}
%commentato per versione da workshop

In this section, we explore the relationship between disinformation flow and the characteristics of accounts in our dataset. We determine the bot scores of the accounts using BotomerLite, Section~\ref{sec:botdetection}. The bot score provides a measure of the extent to which an account exhibits bot-like characteristics, on a scale of 0 to 1. The closer the score is to 1, the more likely it is that the account is a bot.

We perform two different analyses. The first analysis aims to determine whether the bots in our dataset exhibit a discernible political orientation. Specifically, we seek to determine whether accounts within the {\sc Rep} community tend to exhibit more automated behavior than those in the {\sc Rep-Dem-Journ} community.
In our second analysis, we aim to examine two critical aspects: 1. the correlation between the type of accounts and their propensity to generate untrustworthy traffic; and 2. exploring potential correlations between automated accounts and traffic associated with swing states.

%commentato per versione da workshop
For the first analysis, we compare the bot score distributions in {\sc Rep} and {\sc Rep-Dem-Journ} using the Mann-Whitney U~\cite{mann1947} and Kolmogorov-Smirnov~\cite{smirnov1939test} statistical tests.  Both tests are used to determine whether two distributions are different, and if so, in what way.
The bot score distributions were created by keeping the bot score of the account that posted each tweet.

%\begin{table*}[]
%    \centering
%    \begin{tabular}{llrrrrr}
%\toprule
%       dist$_A$ &        dist$_B$ &  KS test (dist$_A$,dist$_B$) &  p-value$_{KS}$ &  MWU test (dist$_A$,dist$_B$) &  MWU test (dist$_B$,dist$_A$)  &  p-value$_{MWU}$ \\
%\midrule
%         {\sc Both} &   {\sc Rep} &     \textcolor{red}{0.02} &         \textcolor{red}{0.0} &        0.49 &        0.51 &          \textcolor{red}{0.0} \\
%         {\sc Both} & {\sc Rep-Dem\_Journ} &     \textcolor{red}{0.05} &         \textcolor{red}{0.0} &        0.53 &        0.47 &          \textcolor{red}{0.0} \\
%         {\sc Rep-Dem\_Journ} & {\sc Rep} &     \textcolor{red}{0.07} &         \textcolor{red}{0.0} & 0.45 & 0.55 & \textcolor{red}{0.0}\\
%\bottomrule
%\end{tabular}
%    \caption{Results of the statistical test regarding the distribution of the bot score in the various communities.} \label{tab:statistical_tests}
%\end{table*}

% commentato per versione da workshop 

\begin{figure}[h]
   \centering
   \caption{Bot scores distributions in both communities (left for each panel), {\sc Rep} (center) and {\sc Rep-Dem-Journ} (right). \label{fig:traffic_bot_score_distribution}}
   %OLD
   %\caption{Distributions of bot scores in both communities (left for each panel), {\sc Rep} (center) and {\sc Rep-Dem-Journ} (right). The green line is the mean, the purple line is the median. \man{The box extends from the first quartile to the third quartile. The whiskers extend from the box to the outermost data point within 1.5 times the interquartile range from the box.}\label{fig:traffic_bot_score_distribution}}
   \includegraphics[width=.90\linewidth]{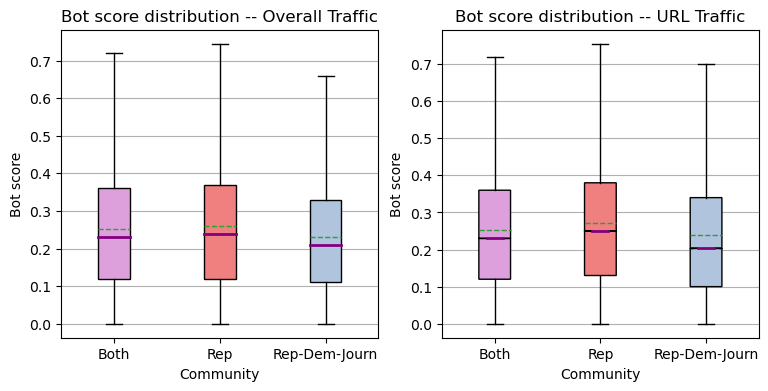}
 \end{figure}

Figure~\ref{fig:traffic_bot_score_distribution} shows the distributions of bot scores with respect to total traffic (left) and traffic containing only URLs (right). Since the distributions associated with the two communities have relatively close means (for the total traffic: {\sc Rep} $0.26$ and {\sc Rep-Dem-Journ} $0.23$; for URL traffic: {\sc Rep} $0.272$ and {\sc Rep-Dem-Journ} $0.238$), we perform the Kolmogorov-Smirnov and Mann-Whitney U statistical tests to assess whether the distributions are statistically different.

The Kolmogorov-Smirnov (KS) test measures the distance between two empirical distributions as the maximum difference in their cumulative distributions. The p-values of the Kolmogorov-Smirnov tests (as shown in Table~\ref{tab:statistical_tests_ks}) indicate that the distributions of bot scores in the two communities are significantly different from that of the entire dataset (p-values are less than $10^{-319}$). 
%Additionally, the bot score distribution of {\sc Rep-Dem-Journ} is slightly farther from that of the entire dataset.
The Mann-Whitney U (MWU) test evaluates the difference in location between the distributions. Our results, as shown in Table~\ref{tab:statistical_tests_mwu}, confirm that the distributions are significantly different (again, all p-values are less than $10^{-309}$). Furthermore, the values of the bot scores in the {\sc Rep} community are higher than those of the entire dataset, and significantly exceed the scores measured in the {\sc Rep-Dem-Journ} community. These results suggest that tweets in the two communities are generated by users with different characteristics in terms of bot scores (Figure~\ref{fig:traffic_bot_score_distribution}).

%commentato per versione da workshop

 \begin{table*}[ht!]
 \caption{Results of the Kolmogorov-Smirnov test about the bot scores distribution in the two main communities.}  \label{tab:statistical_tests_ks}
     \begin{tabular}{llrr}
 \hline
        dist$_A$ &        dist$_B$ &  KS test (dist$_A$,dist$_B$) &  p-value$_{KS}$ \\
 \hline       
         {\sc Validated} &           {\sc Rep} &    0.021 &         $<10^{-319}$\\
         {\sc Validated} & {\sc Rep-Dem-Journ} &    0.047 &        $<10^{-319}$  \\
         {\sc Rep-Dem-Journ} &           {\sc Rep} &    0.068 &          $<10^{-319}$  \\
 \hline
 \end{tabular}
 \end{table*}

%commentato per versione da workshop
 \begin{table*}[ht!]
     \centering
         \caption{Results of the Mann-Whitney U test about the bot scores distribution in the two  communities. }\label{tab:statistical_tests_mwu}
     \resizebox{\textwidth}{!}{\begin{tabular}{llrrr}
 \hline
        dist$_A$ &        dist$_B$ &  MWU test (dist$_A$,dist$_B$) &  MWU test (dist$_B$,dist$_A$)  &  p-value$_{MWU}$ \\
 \hline

         {\sc Validated} &           {\sc Rep}  &    0.486 &    0.514 &           $<10^{-309}$ \\
          {\sc Validated} & {\sc Rep-Dem-Journ} &    0.532 &    0.468 &           $<10^{-309}$ \\
 {\sc Rep-Dem-Journ} &           {\sc Rep}  &    0.453 &    0.547 &           $<10^{-309}$ \\
 \hline
 \end{tabular}}
 \end{table*}

The second analysis aims to detect untrustworthy tweets posted by bot accounts, and the potential correlation between bots and swing-related tweets. 
To identify which accounts are bots, we take the conservative approach used in~\cite{Ferrara2020characterizing}: we classify as bots those accounts `that fall  at the upper end of the bot score distribution'. This approach has the dual benefit of preventing misclassification of accounts with borderline scores, while focusing on accounts with clear bot characteristics.
In practice, we tag each account in the validated dataset using BotometerLite, sort them from lowest to highest bot score, and isolate those with bot scores in the first and last deciles. In the first decile we have real accounts, while in the last decile we have bot accounts. Specifically, the first decile contains accounts with bot scores in the range [0, 0.04], and the last decile contains accounts with bot scores in the range [0.45, 1].
We collect tweets from both real accounts and bots to investigate the source of untrustworthy traffic. We acknowledge that we exclude many accounts from our validated dataset by not including those with bot scores in the range [0.04, 0.45]. Nevertheless, this approach provides us with more reliable guarantees to minimize false positive and false negative predictions.

Table~\ref{tab:human_bot_user_classification} shows statistics for classified accounts. Of the total number of classified accounts, 47.19\% are bots. Considering only the {\sc Rep} community, this percentage increases to 54.93\%, while in the {\sc Rep-Dem-Journ} community it decreases to 37.35\%. In terms of posting activity, bots appear to be more active than real accounts, being about twice as active in both posting tweets and sharing tweets with URLs.
Of the total traffic generated by classified accounts, bots contribute 64.19\% of the traffic, reaching 68.07\% in the {\sc Rep} community and dropping to 53.94\% in the {\sc Rep-Dem-Journ} community. 
\begin{table}
\caption{Genuine and bot accounts in the validated dataset and in the main political communities.\label{tab:human_bot_user_classification}}
\begin{tabular}{lccc}
\hline
        Label &  No. Users &  No. Tweets &  No. URL \\
\hline
        \multicolumn{4}{l}{\sc{Validated dataset}} \\ 
        human &      57797 &      228378 &    25422 \\
        bot &      51648 &      409449 &    53017 \\
        \hline
        \multicolumn{4}{l}{\sc{Rep} }\\ 
         human &      27624 &      147772 &    16156 \\
         bot &      33663 &      315065 &    41383 \\
         \hline
         \multicolumn{4}{l}{\sc{Rep-Dem-Journ}} \\ 
         human &      30173 &       80606 &     9266 \\
         bot &      17985 &       94384 &    11634 \\
\hline
\end{tabular}

\end{table}

%commentato per versione da workshop
\subsubsection{Disinformation, bots, discursive communities, and swing states}\label{sec:BotComm}
Here we focus on the role of bots in spreading links to low-trustworthy/non-trustworthy news stories. Table~\ref{tab:misinformation_for_bot_and_states} shows the percentages of (i) all, (ii) trustworthy (T), and (iii) non-trustworthy (N) URLs shared by users classified as bots or real. The table also takes into account their membership in a discursive community ({\sc Rep} or {\sc Rep-Dem-Journ}) and a state category (swing or safe).

    Focusing on the \emph{Swing \& Safe} column in Table~\ref{tab:misinformation_for_bot_and_states}, we see that about 73\% of the untrustworthy (N) traffic is generated by bots, regardless of the community, 
    while bots are responsible for about $\sim63\%$ of tweets with trustworthy URLs.  
    If we focus only on the traffic generated in {\sc Rep}, the bots spread 73. 84\% of the N and 69.11\% of the T links. 
    Focusing only on untrustworthy links, of the 91\% of the total in swing states, more than 74\% are posted or retweeted by bots. Furthermore, while untrustworthy links associated with safe states are only a small part of the total (8.47\%), the vast majority of this traffic comes from bot accounts (68.75\%).

\begin{table*}[]
%commentata caption per la versione workshop
   % \caption{Percentages of links shared, per reputability\sout{, per community} and per state type. \label{tab:misinformation_for_bot_and_states}}
    \caption{Percentages of links shared, per reputability, per state type and per discursive community\label{tab:misinformation_for_bot_and_states}}

\resizebox{\textwidth}{!}{\begin{tabular}{lcccccccccccc}
\hline
    &&&&& \multicolumn{2}{c}{\emph{\textbf{Swing \& Safe}}} && \multicolumn{2}{c}{\emph{\textbf{Swing}}} && \multicolumn{2}{c}{\emph{\textbf{Safe}}}\\
    %\cmidrule{6-7}
    %\cmidrule{9-10}
    %\cmidrule{12-13}
    
    Community &  No. URL &  swing &  safe &&  bot &  human &&  bot &  human &&  bot &  human \\
    \hline
    %{\sc All Links} & 78439 &  89.92 & 10.08 &&      \textbf{67.59} &        32.41 &&   \textbf{67.87} &     32.13 &&   \textbf{65.06} &     34.94 \\
    %commentatate righe sotto per la versione workshop
    \multicolumn{10}{l}{\textbf{All Links}} \\
    {\sc Validated dataset} & 78439 &  89.92 & 10.08 &&      \textbf{67.59} &        32.41 &&   \textbf{67.87} &     32.13 &&   \textbf{65.06} &     34.94 \\
     {\sc Rep} & 57539 &  90.78 &  9.22 &&      \textbf{71.92} &        28.08 &&   72.26 &     27.74 &&   68.63 &     31.37 \\
    {\sc Rep-Dem-Journ} & 20900 &  87.54 & 12.46 &&      55.67 &       44.33 &&   55.36 &     44.64 &&   57.77 &     42.23 \\
    
    \hline
    %{\sc Trustworthy Links (T)} & 23036 &  83.07 & 16.93 &&      \textbf{62.90} &        37.10 &&   62.69 &     37.31 &&   63.96 &     36.04 \\
    %commentatate righe sotto per la versione workshop
    \multicolumn{10}{l}{\textbf{Trustworthy Links (T)}} \\
         {\sc Validated dataset} & 23036 &  83.07 & 16.93 &&      \textbf{62.90} &        37.10 &&   62.69 &     37.31 &&   63.96 &     36.04 \\
           {\sc Rep} & 11812 &  83.46 & 16.54 & &     \textbf{69.11} &        30.89 &&   68.84 &     31.16 &&   70.47 &     29.53 \\
 {\sc Rep-Dem-Journ} & 11224 &  82.65 & 17.35 & &     56.37 &        43.63 &&   56.15 &     43.85 &&   57.42 &     42.58 \\
    
    \hline
    %{\sc Non-trustworthy Links (N)}& 20627 &  \textbf{91.53} &  \textbf{8.47} & &     \textbf{73.69} &        26.31 &&   \textbf{74.15} &     25.85 &&   \textbf{68.75} &     31.25 \\
    %commentatate righe sotto per la versione workshop
    \multicolumn{10}{l}{\textbf{Non-trustworthy Links (N)}} \\
         {\sc Validated dataset} & 20627 &  \textbf{91.53} &  \textbf{8.47} & &     \textbf{73.69} &        26.31 &&   \textbf{74.15} &     25.85 &&   \textbf{68.75} &     31.25 \\
           {\sc Rep} & 20147 &  91.42 &  8.58 & &     \textbf{73.84} &        26.16 &&   74.33 &     25.67 &&   68.59 &     31.41 \\
 {\sc Rep-Dem-Journ} & 480 &  96.25 &  3.75 &  &    67.50 &        32.50 &&   66.88 &     33.12 &&   83.33 &     16.67 \\
\hline
\end{tabular}}
   
\end{table*}

\section{Discussion}
The study of disinformation within online social networks during election campaigns has yielded a wealth of contributions, exemplified by works such as Becatti et al.~\cite{Becatti2019d}, Bovet and Makse~\cite{Bovet2019influence}, Budak et al.~\cite{budak2019happened}, Ferrara et al.~\cite{Ferrara2020characterizing}, Georgacopoulos et al.~\cite{Georgacopoulos2020how}, Luceri et al.~\cite{luceri2019evolution}, and Mattei et al.~\cite{mattei2022bowtie}, to name a few. However, the spread of untrustworthy content has rarely been linked to the specifics of a particular electoral system; most existing disinformation studies tend to focus on individual countries. Yet emerging evidence suggests that the electoral process plays a role in shaping the dynamics of online discourse. Limited findings to date~\cite{Bright2018,Urman2020,VanVliet2021,Praet2021, howard2018social} suggest that there are differences in how online accounts organize themselves in discussions, either promoting divisive or cohesive structures, depending on whether a country uses majoritarian, proportional, or plurality voting systems.

In our current research, while still focusing on a single country, we direct our attention to two specific aspects: (i) a feature of its presidential electoral system-the presence of swing and safe states-and (ii) whether and to what extent this feature is reflected in the spread of online disinformation.

To elaborate further, each U.S. state is allocated a certain number of presidential electors, and after the statewide popular vote, the faction that receives the highest number of votes claims all of the electors, regardless of the margin of victory. Safe states are those where election outcomes can be easily predicted, while swing states represent fiercely contested battlegrounds that are crucial to securing the presidential election.

With this context in mind, our analysis focuses on the 2020 U.S. presidential election. We focus specifically on the Twitter discourse surrounding eight states, four of which are categorized as safe states (New Jersey, Indiana, Washington, and Louisiana) and the remaining four as swing states (Arizona, Florida, Michigan, and Pennsylvania). We then selected tweets that contained the names of the presidential candidates (either Biden or Trump) and the names of one of the selected states in their text. 

Our first result is that 88\% tweets in our dataset is related to swing states. This underlines the importance of swing states (as opposed to safe) in the political discussion.

Secondly, from Table~\ref{tab:misinformation_in_swing_and_safe} we observe that the frequency of untrustworthy URLs shared in the political debate of swing states (26.06%23.47\%
) is greater than the analogous of safe states (22.25\%%18.33\%
). Symmetrically, the frequency of trustworthy URLs is higher in safe states (50.66\%%50.87\%
) than swing ones (28.30\%%29.84\%
). 
In this sense, not only the debate, but also the spread of disinformation is more intense in swing states due to their importance for the election outcome. To summarize, both the total flow of news and the frequency of untrustworthy URLs are higher in swing states.

Thirdly, we investigate the exposure to disinformation of the two main emergent discursive communities:  
a great community of Republican supporters (the {\sc Rep} community) and a mixed one, including both Democrats and Republicans, as well as various journalists (the {\sc Rep-Dem-Journ} community). Remarkably, the {\sc Rep} community hosts 91\% of the total URLs pointing to  untrustworthy news sources. In addition, each untrustworthy URL in the {\sc Rep} community is shared, on average, more than any other type of URL.

Finally, we investigate the contribution of automated accounts in the spreading of disinformation. Let the reader consider Table~\ref{tab:misinformation_for_bot_and_states}: bots appear to be more active than genuine accounts in posting tweets, both in swing and in safe states, with comparable percentages, i.e. $\sim$67\% vs. $\sim$65\%, respectively in swing and safe states. Regarding the untrustworthy links shared in swing states, more than 74\% are posted or retweeted by bots.

Our analyses were conducted through a careful filtering process applied to the original dataset. We used techniques rooted in Information Theory and Statistical Mechanics principles related to complex networks, as discussed in Section~\ref{sssec:DisCo}, to elucidate political communities. In particular, we focused on the bipartite network representing retweet interactions between verified and unverified users.

To validate the projection of the bipartite network onto the verified user layer, we employed the BiCM (Bipartite Configuration Model) as a benchmark. This involved establishing links between verified users if the number of shared unverified retweeters was statistically significant. We then ran a community detection algorithm on the resulting network of verified users. To extend these communities to unverified Twitter users, we leveraged our knowledge of verified users and implemented a label propagation procedure. Our validation approach ensures that we account for interactions that cannot be attributed solely to user degree sequences, which distinguishes our work from similar studies such as that of Howard et al.~\cite{howard2018social}, who analyzed disinformation flows in swing and safe states during the 2016 election but did not employ entropy-based null models.

In sum, our hypothesis that the spread of disinformation is more pronounced in swing states finds robust support in the data. Due to their pivotal role in determining election outcomes, swing states not only attract a higher volume of tweets, but also bear a greater percentage of the brunt of disinformation campaigns compared to safe states. This disparity in the impact of disinformation, coupled with the increased flow of messages, leads to a particularly worrisome spike in disinformation messages.

\paragraph{Limitations and future work}
While our findings provide compelling insights, it is important to acknowledge certain limitations that invite further investigation. First, our research was limited to a select number of U.S. swing and safe states, providing a specific snapshot of the broader electoral landscape. In addition, our analysis was limited to the 2020 U.S. presidential election. Expanding our study to include comparative analyses with the 2012 and 2016 elections could either validate our conclusions or contextualize them in the context of the 2020 contest.

\man{We also focus on a subset of swing and safe states, rather than all of them. Our choice was a compromise between a number of practical limitations. First, we needed a dataset of manageable size, so we limited our analyses to a subset of swing states, focusing on the four largest. Second, we needed an appropriate benchmark, i.e., a sample of safe states against which to compare our measurements. Such a choice was more complicated because the number of safe Republican states tends to be much larger than the number of safe Democratic states, but they tend to be less populous and thus represent a smaller number of electoral votes. In this sense, a good compromise was to choose four safe states, i.e. the same number of swing states, equally divided between Democrats and Republicans, with almost the same number of electoral votes. 
In addition, it is important to note that even if the states we chose to study happen to exhibit a flow of misinformation that is significantly different from that of other solid states, we still have evidence of an increased focus by misinformation producers on states that are more likely to influence the final outcome of the national election.}

Furthermore, our data collection methodology relied on keyword-based approaches that inherently lack a precise understanding of the exact content of the collected tweets. Although the presence of both state and candidate names in the tweets implies a connection to the election and the state, the specific content remains unknown until examined.

Extending our study to other plurality voting systems with similarities to the U.S., such as the United Kingdom, would provide valuable insights into the presence of analogous disinformation diffusion dynamics within swing constituencies. In addition, examining the presence of disinformation at the geographic level within different electoral systems, including proportional systems (e.g., Germany and Spain), majoritarian systems (e.g., France), or mixed systems (e.g., Italy, South Korea, and Japan), would further enrich our understanding of this phenomenon.

Finally, we argue that our research contributes to a more detailed examination of the relationship between electoral systems, online discourse, and the spread of online disinformation.

\section*{Declarations}

\subsection*{Availability of data and materials}
The dataset supporting the conclusions of this article is available in the dataverse repository, \url{https://doi.org/10.7910/DVN/ANBPTC}. The data about the reliability of the various news sources comes from NewGguard, but restrictions to their availability apply since they were used under a NewsGuard license and they are not publicly available. These data could be however available upon reasonable request and with permission of Newsguard. %{\bf We are pleased to share our dataset with the scientific community, hoping to foster further exploration and deeper insights into this important issue.}
 
\subsection*{Competing interests}
The authors declare that they have no competing interests.

\subsection*{Funding}
This work is partially supported by project SERICS (PE00000014) under the MUR National Recovery and Resilience Plan funded by the European Union - \#NextGenerationEU, by the Integrated Activity Project TOFFEe (TOols for Fighting FakEs) \url{https://toffee.imtlucca.it/} and by the IIT-CNR funded Project re-DESIRE (DissEmination of ScIentific REsults 2.0).

\subsection*{Author's contributions}
The authors declare equal contribution.

\subsection*{Acknowledgements}
The authors thank the reviewers for their valuable suggestions.

%%%%%%%%%%%%%%%%%%%%%%%%%%%%%%%%%%%%%%%%%%%%%%
%%                                          %%
%% Backmatter begins here                   %%
%%                                          %%
%%%%%%%%%%%%%%%%%%%%%%%%%%%%%%%%%%%%%%%%%%%%%%

\begin{backmatter}

%%  After submission of the .TEX file,                     %%
%%  you will be prompted to submit your .BBL file.         %%

% if your bibliography is in bibtex format, use those commands:
\bibliographystyle{bmc-mathphys} % Style BST file (bmc-mathphys, vancouver, spbasic).
\bibliography{bmc_article}      % Bibliography file (usually '*.bib' )

\subsection*{Figure Legends and Table Legends}
\begin{itemize}
    \item  Table \ref{table:domains-tags}: Tags are inherited from NewsGuard, the UNC tag indicates that NewsGuard has not yet tagged that domain. 
    \item Table \ref{tab:community_characterization}: Two main communities emerge, {\sc Rep} and {\sc Rep-Dem-Journ}. With {\sc Others} we characterize all the accounts that do not belong to the giant component: their contribution is going to be disregarded in the following, since they do not contribute  to the entire debate.
    \item Figure \ref{fig:retweet_network}: The two main communities that emerge are the red one, characterized by Republican supporters, and the blue one, a mix of Republicans, Democrats, and journalists of various affiliations. 
    \item Figure \ref{fig:url_reliability_distribution}: trustworthy news publisher (T); untrustworthy news publisher (N); not a news site, e.g. platforms as amazon.com (P); link corresponding to a domain not classified by NewsGuard (UNC).
    \item Figure \ref{fig:sharing_by_reliability}: Values are normalized applying a log function. The green line is the mean, the purple line is the median; outliers are not reported. \man{The box extends from the first quartile to the third quartile. The whiskers extend from the box to the outermost data point within 1.5 times the interquartile range from the box.}
    \item Figure \ref{fig:url_sharing_behaviour_in_swing_and_comm}: Values are normalized applying a log function. The green line is the mean, the purple line is the median.\man{The box extends from the first quartile to the third quartile. The whiskers extend from the box to the outermost data point within 1.5 times the interquartile range from the box.}
    \item Figure \ref{fig:traffic_bot_score_distribution}: The green line is the mean, the purple line is the median. \man{The box extends from the first quartile to the third quartile. The whiskers extend from the box to the outermost data point within 1.5 times the interquartile range from the box.}
\end{itemize}

\subsection*{List of abbreviations}

\begin{table}[ht!]

\begin{tabular}{ll}
\hline
Abbreviation & Definition\\
\hline
        API & Application Programming Interface \\ 
        BiCM & Bipartite Configuration Model \\ 
        MEP & Member of the European Parliament \\ 
        MWU & Mann-Whitney U test \\ 
        N & Untrustworthy news domain \\ 
        No. & Number \\ 
        P & Platform domain (e.g., reddit.com, twitter.com) \\ 
        QAnon & American political conspiracy theory and political movement \\ 
        {\sc Rep-Dem-Journ} & Community of mixed users, comprising journalists as well as both Republican and Democratic supporters \\ 
        {\sc Rep} & Community of Republican supporters \\ 
        S & Satire domain \\ 
        T & Trustworthy news domain \\ 
        U.S. & United States \\ 
        UNC & Unclassified domain \\ 
        URL & Uniform Resource Locator \\ 
        VIP & Very Important People \\ 
\hline
\end{tabular}
\end{table}

\end{backmatter}
\end{document}